\documentclass[a4paper,11pt]{article}
\usepackage{subfiles}
\pdfoutput=1 % if your are submitting a pdflatex (i.e. if you have
\usepackage{amsmath,amsfonts,amssymb,graphicx,color}
\usepackage{longtable,booktabs}
\usepackage{placeins}

\AtBeginDocument{
\heavyrulewidth=.08em
\lightrulewidth=.05em
\cmidrulewidth=.03em
\belowrulesep=.65ex
\belowbottomsep=0pt
\aboverulesep=.4ex
\abovetopsep=0pt
\cmidrulesep=\doublerulesep
\cmidrulekern=.5em
\defaultaddspace=.5em
}

\usepackage{geometry}
\usepackage{setspace}
\usepackage{changepage}  
\usepackage{xspace}
\usepackage{jcappub} % for details on the use of the package, please
\usepackage{orcidlink}
\usepackage[T1]{fontenc} % if needed
                     
\usepackage{multirow}      
\usepackage[bf]{caption}

\usepackage{subcaption}

\usepackage{blkarray, bigstrut}
\usepackage{xparse}

\usepackage{lipsum}  %joint figures
\usepackage{graphbox}   % allows to add keys to \includegraphics
\usepackage{stackengine}
\usepackage{graphicx}% Include figure files
\usepackage{dcolumn}% Align table columns on decimal point
\usepackage{hyperref}
\hypersetup{colorlinks}
\usepackage[T1]{fontenc} % if needed
\usepackage[normalem]{ulem}
\usepackage[utf8]{inputenc}

\usepackage[nameinlink,noabbrev]{cleveref}
\crefname{section}{Section}{Sections}
\crefname{table}{Table}{Tables}
\crefname{appendix}{Appendix}{Appendices}
\Crefname{figure}{Figure}{Figures}
\Crefname{equation}{Equation}{Equations}
\Crefname{section}{Section}{Sections}
\Crefname{table}{Table}{Tables}

\usepackage{booktabs}
\usepackage{float} % para usar[H]

\usepackage{soul}

\usepackage{bm}
\let\vec\bm

\newcommand{\mJ}{\mathcal{J}}

\newcommand{\ve}[1]{{\text{\bf #1}}} 

\newcommand{\vd}{\vec d}

\newcommand{\vk}{\vec k}
\newcommand{\vp}{\vec p}

\newcommand{\vx}{\vec x}

\newcommand{\vhn}{\hat{\vec n}}

\newcommand{\dD}{\delta_\text{D}}

\newcommand{\vt}{\boldsymbol\theta}
\newcommand{\vell}{\boldsymbol\ell}

\newcommand{\model}{\boldsymbol\mu}

\newcommand{\vnu}{\boldsymbol\nu}

\newcommand{\aaa}[1]{\textcolor{red}{#1}}

\usepackage{float} % para usar [H]

\usepackage{booktabs}
\title{Non-Gaussian statistics in galaxy weak lensing: compressed three-point correlations and cosmological forecasts}

\author[a]{{Sofia Samario-Nava}\orcidlink{0009-0001-5224-9153},}
\emailAdd{ssamario@icf.unam.mx}

\author[a]{{Alejandro Aviles}\orcidlink{0000-0001-5998-3986},}
\emailAdd{aviles@icf.unam.mx}

\author[a]{{Juan Carlos Hidalgo}\orcidlink{0000-0001-9715-1232}}
\emailAdd{hidalgo@icf.unam.mx}

\affiliation[a]{Instituto de Ciencias F\'isicas, Universidad Nacional
Autónoma de México,  62210, Cuernavaca, Morelos.}

\keywords{Large-scale structure, weak lensing higher-order statistics, Fisher Forecasts}

\abstract{
Building on previous developments of a harmonic decomposition framework for computing the three-point correlation function (3PCF) of projected scalar fields over the sky, this work investigates how much cosmological information is contained in these higher-order statistics. We perform a forecast to determine the number of harmonic multipoles required to capture the full information content of the 3PCF in the context of galaxy weak lensing, finding that only the first few multipoles are sufficient to capture the additional cosmological information provided by the 3PCF. This study addresses a critical practical question: to what extent can the high-dimensional 3PCF signal be compressed without significant loss of cosmological information? Since the different multipoles contain highly redundant information, we apply a principal component analysis (PCA) which further reduces its dimensionality and preserving  information. We also account for non-linear parameter degeneracies using the DALI method, an extension of Fisher forecasting that includes higher-order likelihood information. %Under optimistic settings, we find that the 3PCF improves considerably the constraining power of the 2PCF for the matter abundance $\Omega_m$ and fluctuations amplitude $\sigma_8$ by about 18\% for galaxy maps at redshift $z=0.5$. However, for the amplitude $S_8$ and the equation of state parameter of dark energy, $w_0$, the improvements are very modest. Thanks to the harmonic basis and PCA, the method can be implemented with relatively low computational cost, providing a practical tool for current and future photometric surveys.
Under optimistic settings, we find that the 3PCF considerably improves the constraining power of the 2PCF for $\Omega_m$, reaching a 20\% improvement. Other parameters also benefit, mainly due to their degeneracy with the matter abundance. For example, with our chosen scale cuts for galaxy sources at $z = 0.5$, we find that $\sigma_8$ is more tightly constrained, whereas $S_8$ and $w_0$ are not.
Finally, we construct analytical Gaussian covariance matrices that can serve as a first step toward developing semi-analytical, semi-empirical alternatives to sample covariances.
}

\begin{document} 
\maketitle
\flushbottom

\section{Introduction}

In the standard, dominant picture of our Universe, an early inflationary period is responsible for amplifying primordial quantum fluctuations, leading to adiabatic, nearly Gaussian, and scale-invariant initial conditions on the cosmic field perturbations \cite{Guth1981,Linde1982,MukhanovChibisov1981,Bartolo:2004if}. These initial conditions undergo gravitational collapse, giving rise to the large-scale structures (LSS) we observe today \cite{Peebles1980,Springel2005}. In the early stages of the Universe, field fluctuations remain small, and linear gravity theory preserves their Gaussian profile. Consequently, cosmic microwave background (CMB) measurements tightly constrain primordial non-Gaussianity, with no significant detection to date \cite{DESI:2023duv,Bermejo-Climent:2024bcb,Jung:2025nss}.
However, the LSS is non-Gaussian due to non-linear gravitational collapse, which (i) causes a redistribution of statistical information from two-point functions to higher-order statistics \cite{Bernardeau:2001qr}, and (ii) imprints signatures of non-linear dynamics  \cite{Scoccimarro:2000sn}.
Higher-order statistics thus complement traditional 2-point analyses by capturing this leakage of information and probing deviations from standard scenarios, such as in modified gravity \cite{Bose:2019wuz,Munshi:2020nfh,Cataneo:2021xlx,Aviles:2023fqx,Hou:2023kfp,Sugiyama:2023tes} or the free-streaming of massive neutrinos \cite{deBelsunce:2018xtd,Uhlemann:2019gni,Kamalinejad:2020izi,Aviles:2021que,Pal:2025hpl}.

Unfortunately, the extraction of higher-order statistics from catalogs can be a difficult task for several reasons. The immediate challenge is that, at large scales, such signals are expected to be small. Furthermore, they are often buried under cosmic variance and shot noise, requiring large survey volumes and precise control of systematics to detect them reliably. Also, the computation of higher-order correlators from the data (e.g., bispectra, trispectra) is significantly more demanding than 2-point statistics, both computationally and in terms of their sensitivity to observational artifacts, including survey geometry and incomplete sky coverage. Computationally, for $N$-point functions, the total number of calculations scales with the number of points $N_p$ in a sample as $\mathcal{O}(N_p^N)$, or  $\mathcal{O}(N_p^{N} \log N_p)$ when correlations fall out rapidly, and one is able to use tree search algorithms. 

A particularly and increasingly powerful observable in this context is the convergence field $\kappa$ derived from galaxy weak gravitational lensing measurements. The convergence  $\kappa$  quantifies the integrated matter distribution along the line of sight between the observer and the image, effectively projecting the 3D matter density field onto a 2D map \cite{Bartelmann2001}. This makes $\kappa$ a direct tracer of the total matter (dark and baryonic), with the potential of playing an important role for probing structure growth and the cosmic expansion, then providing promising summary statistics to test modified gravity, dark energy, and massive neutrinos, among other science cases \cite{Kilbinger2015,Mandelbaum2018}.
Unlike other observables over the sky, as CMB anisotropies,  $\kappa$ is inherently non-Gaussian due to non-linear structure formation and projection effects across redshifts. Thus, two-point statistics alone (e.g., its power spectra) cannot capture its full cosmological information. Higher-order statistics, such as bispectra (or the 3PCF), peak counts, Minkowski functionals, the (1D) probability distribution function, among others, might be invaluable for breaking parameter degeneracies and accessing non-Gaussian regimes \cite{Zaldarriaga:2002qt,Schneider:2002ze,Takada:2003sv,Kratochvil:2011eh,Petri:2013ffb,Liu:2016nfs,DES:2017hhj,Halder:2021itp,Uhlemann:2022znd,Barthelemy:2023mer,Euclid:2023uha,2025arXiv250303964G}.
Current surveys like the Dark Energy Survey (DES), the Kilo Degree Survey (KiDS) and the Hyper Suprime-Cam (HSC) have already begun constraining these statistics, though limitations in sky coverage and depth restrict their sensitivity to mildly non-linear scales \cite{DES:2021vln,DES:2022oqz,Burger:2023qef,Marques:2023bnr,Thiele:2023gqr}. In contrast, the Legacy Survey of Space and Time (LSST) of the Vera Rubin Observatory, will provide high-precision measurements on almost half of the sky, significantly reducing sample variance and shot noise \cite{LSSTScience:2009jmu,LSSTDarkEnergyScience:2018jkl,LSSTDarkEnergyScience:2020oya}. However, extracting these signals requires overcoming key challenges: convergence reconstruction amplifies uncertainties from intrinsic galaxy shapes, photometric redshift errors, and shear calibration biases (e.g. \cite{Troxel:2014dba}), while theoretical modeling must account for non-linear gravity, baryonic effects, and survey systematics \cite{Semboloni:2011fe,Leauthaud:2016jdb}.

Building on previous works \cite{Arvizu:2024rlt,Porth:2023dzx,Sugiyama:2024uqo}, which introduced an efficient harmonic-space framework for computing the three-point correlation function (3PCF) of projected fields, and drawing from earlier methods in 3D galaxy counts \cite{Szapudi:2004gg,Zheng:2004eh,Pan:2005ym,Slepian:2015qza,Slepian:2016weg,Slepian:2016kfz,Philcox:2021bwo,Hou:2021ncj}, in this work we quantify the cosmological information content of these higher-order statistics. Through cosmological forecasts, based on the Fisher information matrix and non-Gaussian extensions, we determine how many harmonic moments of the 3PCF are needed to saturate its information. This addresses a critical question for upcoming surveys like LSST: Given computational constraints, to what extent can we compress the 3PCF without significant information loss? Our analysis might be useful for future DES Year 6 \cite{DES:2025key}, Euclid \cite{EUCLID:2011zbd,Euclid:2023uha} and LSST analyses, identifying which harmonic modes contribute most to cosmological constraints. 

The paper is organized as follows: In \cref{Modeling}, we present the theoretical modeling and introduce the \texttt{3pt-WL} code. In \cref{sec:data}, we review the data used in this work, which was obtained in \cite{Arvizu:2024rlt}. In \cref{sec:modelvsdata}, we compare the model predictions with the data, use this comparison to define our data vector, and construct the sample covariance matrices. In \cref{sec:Forecast1pole}, we compute Fisher forecasts for a single multipole and apply the DALI method. Our main results are presented in \cref{sec:PCA}, where we compress the data vector using PCA and show cosmological forecasts for the combined 2PCF and 3PCF multipoles. In \cref{sec:GaussianCov}, we develop an analytical model for the covariance matrix, with detailed calculations provided in \cref{app:cov}. Finally, in \cref{sec:conclusions}, we present our conclusions.

\section{Modeling the 3PCF}\label{Modeling}

Light rays emitted from distant galaxies are subtly distorted by the intervening matter distribution between them and us. Although these displacements correspond to very small angles on the sky, typically sub-arcseconds, they are still significant enough to produce percent-level changes in the observed ellipticities of galaxies. By averaging these shape distortions over large numbers of galaxies, we can extract valuable cosmological information about the background evolution and growth of structure in the Universe. In the weak-lensing regime, the relationship between the observed and intrinsic galaxy images is described by a linear transformation characterized by the weak lensing convergence $\kappa$ and the shear components $(\gamma_1, \gamma_2)$. These quantities are not independent, but they are related through the lensing potential $\psi_L$, which represents a weighted projection of the gravitational potentials along the line of sight. 
% The relations are given by, see, e.g. \cite{Kilbinger2015},
% \begin{align*}
%     &\kappa(\vt) = -\frac{1}{2} (\partial_1 \partial_1 + \partial_2\partial_2 ) \psi(\vt) = -\frac{1}{2} \nabla^2 \psi(\vt) \\
%     &\gamma_1(\vt) = -\frac{1}{2} (\partial_1 \partial_1 -  \partial_2\partial_2 ) \psi(\vt),\quad 
%     \gamma_2(\vt) = -\partial_1 \partial_2 \psi(\vt), 
% \end{align*}
% using the first equation in the form  \tct{$\psi(\vt) = - 2 \nabla^{-2} \kappa(\vt)$}, we have
% \begin{align}
% \gamma(\vt) =  (\partial_1 + i  \partial_2)^2 \, \nabla^{-2} \kappa(\vt), 
% \end{align}
% \begin{align*}
% \gamma(\vt) %&=  (\partial_1 + i  \partial_2)^2 \, \nabla^{-2} \kappa(\mt) \\
%             &= \int d^2 \theta' \mathcal{D}(\vt-\vt') \kappa(\vt'),\quad \text{with} \quad 
%             \mathcal{D}(\vt) = \frac{\theta_2^2-\theta_1^2 - 2 i \theta_1\theta_2 }{\theta^4} \\
% \kappa(\vt) %&=  (\partial_1 + i  \partial_2)^2 \, \nabla^{-2} \kappa(\mt) \\
%             &= \int d^2 \theta' \mathcal{D}^*(\vt-\vt') \gamma(\vt')            
% \end{align*} 
%
The convergence quantifies the isotropic magnification (or focusing) of galaxy images, while the shear the anisotropic stretching (or squeezing). In many works of weak lensing higher-order statistics, the convergence field is preferred over the shear due to its scalar nature, which facilitates theoretical investigations into the non-Gaussian features of the lensing signal and their connection to the underlying matter distribution. However, the shear is the directly observed quantity and the convergence field is reconstructed from it through non-local inversion methods \cite{Kaiser:1992ps,Seitz:1995dq,Bartelmann:1995yq,Leonard:2013hia}, which can amplify noise, introduce artifacts, and become particularly challenging in the presence of survey masks or incomplete data (e.g., \cite{Kilbinger2015}).

The convergence $\kappa(\vt)$ from a population of source galaxies with number density distribution $W_g(\chi)$ (as a function of comoving distance $\chi$) is given by a weighted line-of-sight projection of the three-dimensional matter overdensity field $\delta(\chi \vt, \chi)$ along our past light cone,
\begin{align*}
\kappa(\vt) = \int_0^{\chi_\text{lim}} d\chi' q(\chi') \delta(\chi'\vt, \chi'),   
\end{align*}
where $\chi_\text{lim}$ is the limiting comoving distance of the galaxy sample. The lensing efficiency function $q(\chi)$ provides the geometric weighting. Assuming flat space, it can be written as
\begin{displaymath}
q(\chi) = \frac{3 \Omega_m H^2_0}{2 c^2}  \frac{\chi}{a(\chi)} \int_\chi^{\chi_\text{lim}} d \tilde{\chi} \, W_g(\tilde{\chi}) \frac{\tilde{\chi}-\chi}{ \tilde{\chi}}, 
\end{displaymath}

The convergence, hence, becomes a linear measure of the total matter density, projected along the line of sight and weighted by the source galaxy distribution $W_g(\chi)d\chi$.
The simplest statistic one can construct out of the convergence is the 2PCF, which within the Limber approximation \cite{1953ApJ...117..134L,Kaiser:1991qi} is given by
\begin{equation} \label{eq:xi}
    \xi(\theta) = \langle \kappa(\vnu) \kappa(\vnu +\vt) \rangle=\int \frac{d\ell}{2\pi} J_0(\theta \ell) C_\ell(\ell)
\end{equation}
with $C_\ell(\ell)$ the weak lensing convergence angular power spectrum. We have neglected $B$-modes, which are not produced by gravitational collapse. In terms of the dark matter power spectrum, $P_\delta(k;z)$, one can write
\begin{equation} \label{Cell}
    C_\ell(\ell) = \int_0^{\chi_\text{lim}} d\chi \frac{q^2(\chi)}{\chi^2} P_\delta\left(\frac{\ell}{\chi};z(\chi) \right).
\end{equation} 
The 2PCF containts all the statistical information of the $\kappa$ field when its underlying distribution is Gaussian. 
While this can be a good approximation early times, non-linear gravitational evolution induces significant non-Gaussian features. The most straightforward generalization accounting for these contributions is the convergence bispectrum,
\begin{align} \label{Bkappa}
B(\vell_1,\vell_2,\vell_3) &= \int_0^{\chi_\text{lim}} d\chi \frac{q^3(\chi)}{\chi^4} B_\delta\left(\frac{\vell_1}{\chi},\frac{\vell_2}{\chi},\frac{\vell_3}{\chi};\chi \right).
\end{align}
with $B_\delta(\vk_1,\vk_2,\vk_3;\chi)$ the matter bispectrum at redshift $z(\chi)$.

The configuration space counterpart of the bispectrum is the 3PCF, given by the connected ($c$) correlator
\begin{equation}
\label{Ec: Estimator}
\zeta(\vt_1,\vt_2) = \langle(\kappa(\vnu)\kappa(\vnu+\vt_1)\kappa(\vnu+\vt_2)\rangle_c, 
\end{equation}

For a statistical homogeneous and isotropic sample $\kappa$ defined over the sky, we can characterize its connected 3PCF,
by three angles $(\theta_1, \theta_2, \theta_3 = |\vt_2-\vt_1|)$, corresponding to three sides of the triangles connecting the points with vector positions $\vnu$, $\vnu+\vt_1$ and $\vnu+\vt_2$. Alternatively, we can characterize $\zeta$ by two of their sides and the angle  between them. That is,  $\zeta = \zeta(\theta_1,\theta_2,\phi_{21})$, with $\phi_{21}=\phi_2-\phi_1$ the difference between the angles that $\vt_1$ and $\vt_2$ spans with respect to an arbitrary direction $\mathbf{\hat{n}}$, fixed at each pivot point on the sphere. Around that point, we assume the planar approximation and compute the 3PCF multipoles
\begin{equation}
    \zeta_m(\theta_1,\theta_2) = \int_0^{2\pi} \frac{d\phi}{2\pi} e^{\text{i} m\phi} \zeta(\theta_1,\theta_2,\phi), 
\end{equation}
and we can write
\begin{equation}
 \zeta(\theta_1,\theta_2,\phi) = \sum_{m=-\infty}^{\infty} \zeta_m(\theta_1,\theta_2) \cos(m \phi),      
\end{equation}
where we have assumed that the 3PCF is (i) real, such that $\zeta_{-m}=\zeta_{m}^*$, and (ii) it is parity even, such that $\zeta(\theta_1,\theta_2,-\phi)=\zeta(\theta_1,\theta_2,\phi)$. Departures from these assumptions are discussed in \cite{Arvizu:2024rlt}. 

The multipoles (or moments) $\zeta_m$ are the main subject of this work. They are components of the whole 3PCF, but they are also summary statistics in their own right, capturing distinct angular features of the signal and allowing for efficient data analysis. For the analytical modeling, it is useful to relate these multipoles to their Fourier counterparts, through the double Hankel transform
\begin{align} \label{BnToZn}
    \zeta_{m}(\theta_{1},\theta_{2}) &= (-1)^{m} \int_0^\infty \frac{\ell_{1}d\ell_{1}}{2\pi} \int_0^\infty \frac{\ell_{2}d\ell_{2}}{2\pi} J_{m} (\ell_{1} \theta_{1} ) J_{m}( \ell_{2} \theta_{2} )  B_{m}(\ell_1,\ell_2),
\end{align}
where the moments of the bispectrum $B$ are 
\begin{align}
   B_m(\ell_1,\ell_2) &= \int_0^{2\pi} \frac{d\psi'}{2\pi} e^{\text{i} m\psi'} B(\vell'_1, \vell'_2)  \\
   &=  \int_0^{\chi_\text{lim}} d\chi \frac{q^3(\chi)}{\chi^4} \int_0^{2\pi} \frac{d\psi}{2\pi} e^{\text{i} m\psi} B_\delta\left( \frac{\ell_1}{\chi},\frac{\ell_2}{\chi}, \psi; \chi\right), 
\end{align}
with $\psi$ the angle between wavevectors $\vell_1$ and $\vell_2$. 

For the purpose of this work we compute $B_\delta$ using the Bihalofit fitting formulae of \cite{Takahashi:2019hth}, and compute \cref{BnToZn} using the 2D-FFTLog method presented in \cite{Fang:2020vhc}.\footnote{Routines publicly available at \url{https://github.com/xfangcosmo/2DFFTLog}.} 
Together with this paper we release \texttt{3pts-WL},\footnote{\url{https://github.com/alejandroaviles/3pts-WL}} a \texttt{C}-language code that computes the multipoles of the 3PCF of the weak lensing convergence. Despite in this work we only utilize the halo model to compute the matter bispectrum, our code also allows for a perturbation theory/effective field theory implementation developed in \cite{Arvizu:2024rlt}.

\section{Simulated data}\label{sec:data}

In this work, we use the full-sky gravitational lensing simulations from \cite{Takahashi:2017hjr}.\footnote{\url{https://cosmo.phys.hirosaki-u.ac.jp/takahasi/allsky_raytracing/}} These simulations employ multiple-lens plane ray-tracing through high-resolution $N$-body runs, generating $N_\text{sims}=108$ independent full-sky realizations of lensing observables, including galaxy lensing convergence and shear maps, and CMB lensing deflection. The source redshifts cover up to $z = 5.3$, with different bins spaced at intervals of redshift $\Delta z=0.05$, corresponding to $150\,h^{-1}\mathrm{Mpc}$ in comoving distance.  In this work, we use the maps at $z=0.5$, $1.0$ and $2.0$.

Each $N$-body simulation evolves $2048^3$ particles in a box with size sides of $4500\,h^{-1}\text{Mpc}$, resolving dark matter halos down to $\sim$$10^{13}\,h^{-1}M_{\odot}$, sufficient for modeling the large-scale structure relevant to galaxy weak lensing. The ray-tracing incorporates post-Born corrections, improving accuracy beyond the Born approximation. The simulations assume a flat $\Lambda$CDM cosmology consistent with WMAP 9-year parameters: $\Omega_c = 0.233$, $\Omega_b = 0.046$, $h = 0.7$, $n_s = 0.97$,  $\sigma_8 = 0.82$ and $w_0=-1$ , with $\Omega_{c,b}$ the present density of cold dark matter and baryons in units of the critical density, $h=H_0/(100\, \text{km} \,\text{s}^{-1}\text{Mpc}^{-1})$ the reduced Hubble constant, $\sigma_8$ the amplitude of present day perturbations averaged over spheres of radii $8\, h^{-1} \text{Mpc}$, and $n_s$ the spectral index of primordial perturbations.  The equation of state parameter $w_0=P_{de}/\rho_{de} c^2$ is assumed constant and equal to $-1$.

The weak lensing convergence 2PCF and 3PCF  multipoles are computed using the code \texttt{cTreeBalls}\footnote{\url{https://github.com/rodriguezmeza/cTreeBalls}} (or \texttt{cBalls} for short), an efficient C-based numerical tool for calculating correlation functions of projected scalar fields on the unit sphere. The code employs OpenMP for parallelization and defaults to an oct-tree data structure \cite{1986Natur.324..446B} for vertex pair searches, achieving approximately a $\mathcal{O}(N \log N)$ scaling with the number of particles, or HEALPix pixels. kd-tree search algorithms \cite{Bentley:1975:MBS:361002.361007} are also supported, which offer comparable efficiency. For a HEALPix resolution parameter $N_\text{side}=4096$ and 9 multipoles, the typical wall-clock runtime is approximately 10 minutes when utilizing 128 threads on a Perlmutter-NERSC node.\footnote{\url{https://docs.nersc.gov/systems/perlmutter/architecture/}}  The scaling with the number of multipoles is small, growing only in about 20\% when going from computing 1 to 9 multipoles.

In \cite{Arvizu:2024rlt}, we measured the 2PCF and 3PCF multipoles from $m=0$ to 8, in 20 logarithmic-spaced bins over the range $\theta \in [8,200]$ arcmin, where values correspond to bin edges rather than centers. These measurements are utilized throughout this work.

\section{Comparison between simulations and modeling}\label{sec:modelvsdata}

\begin{figure*}
	\begin{center}
	\includegraphics[width=3.5 in]{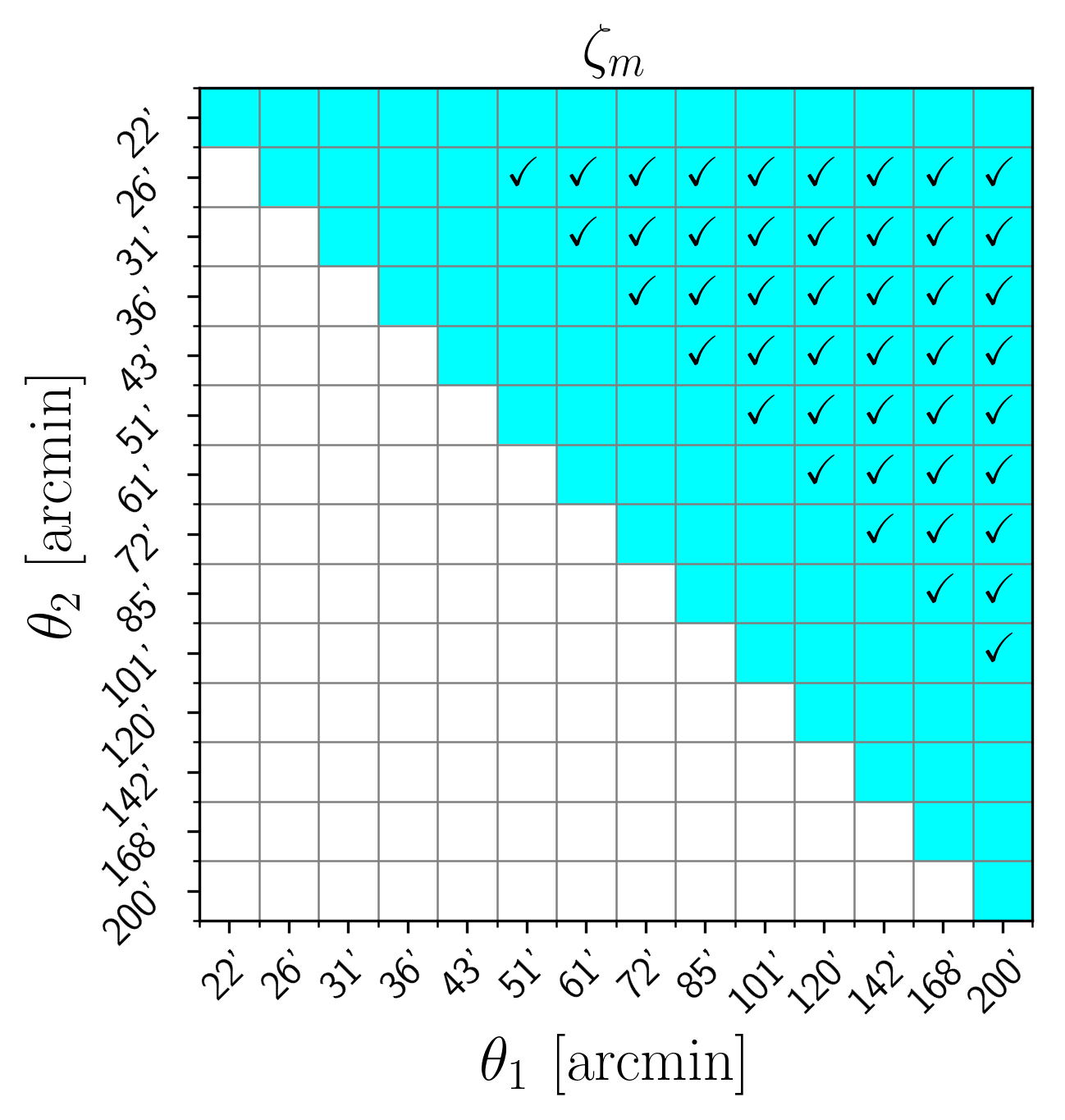}
 \caption{$\zeta_m(\theta_1,\theta_2)$ bins used in this work for a single multipole. The space marked with a checkmark show the bins for which our analytical modeling is within $1\sigma$ of the simulated data in the cases $m=0,1,2,3$.} \label{fig:zetam_bins}
	\end{center}
\end{figure*}

\begin{figure*}
	\begin{center}
	\includegraphics[width=5 in]{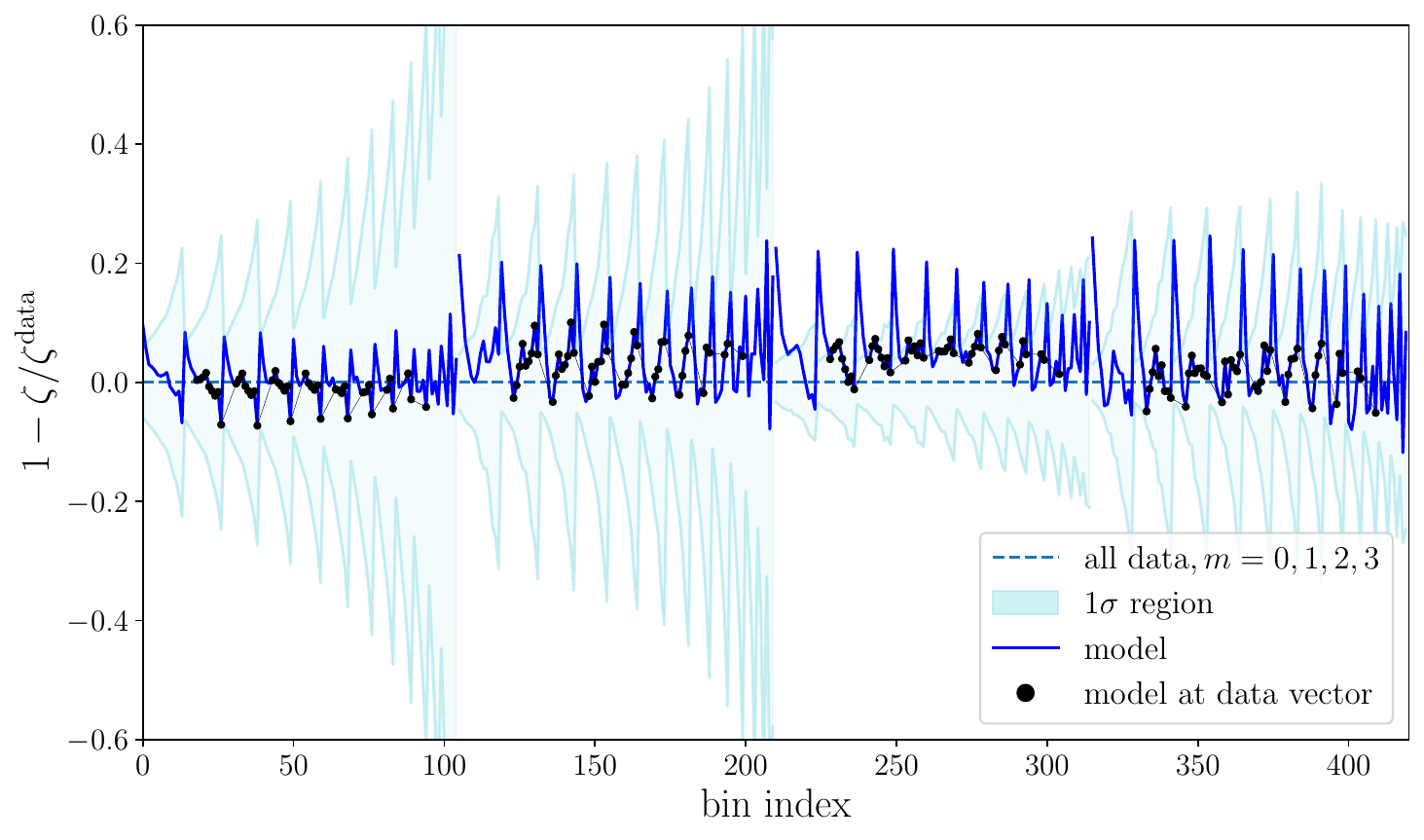}
    \includegraphics[width=5 in]{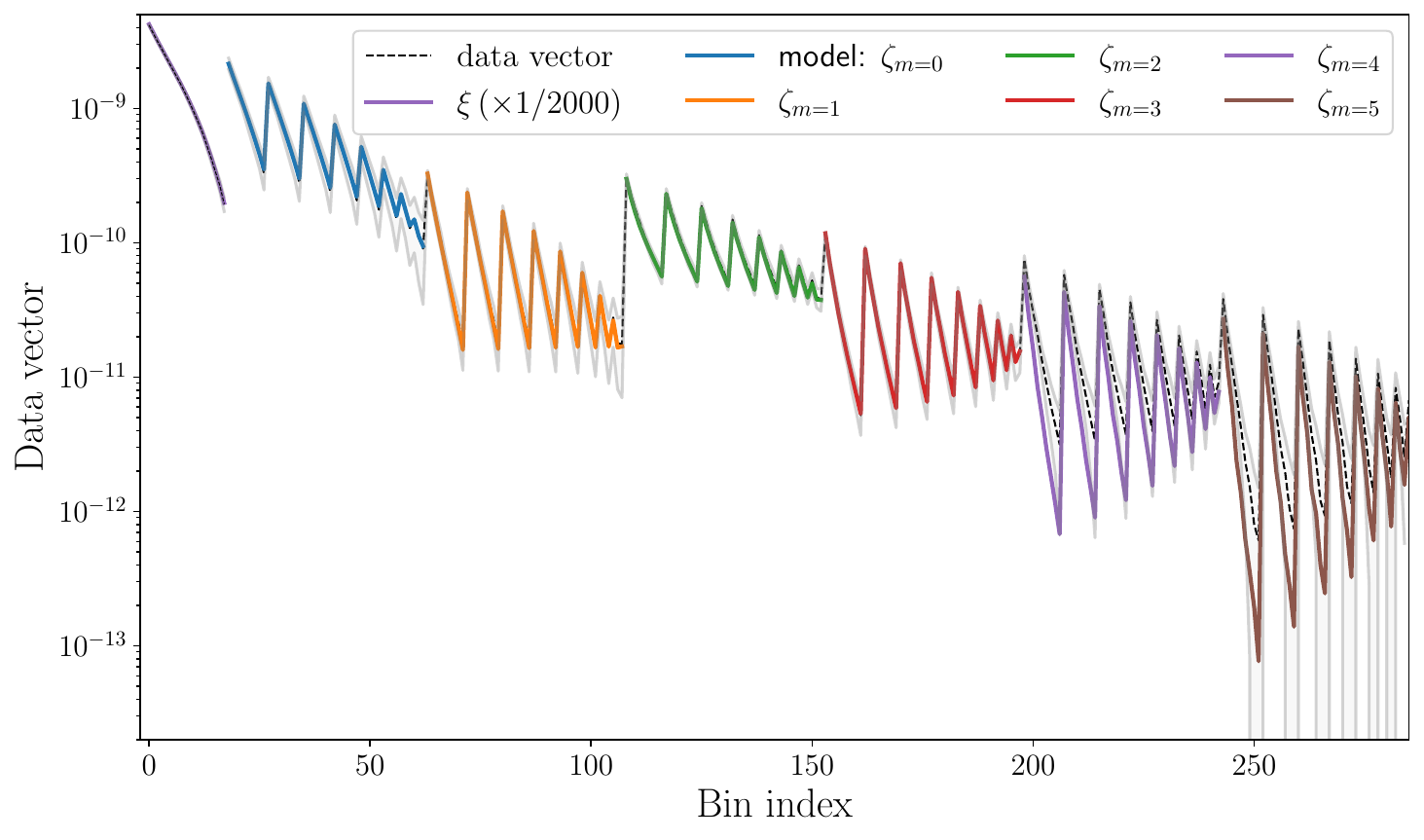}
 \caption{Top panel: Comparison of model and data vectors. The blue lines show the relative difference between the two, while the shaded region indicates the 1$\sigma$ uncertainty derived from the simulated data. The final model vector (black curves) is constructed using the bins marked with a check in \cref{fig:zetam_bins}. Bottom panel: We show only the data vector used  (black dashed lines) and compared to the model vector (solid lines). However, we extend the multipoles up to $m=5$. The shadowing shows the 1$\sigma$ errors. For visualization purposes,  we have  multiplied $\xi$ by $1/2000$. } \label{fig:datavector_all}
	\end{center}
\end{figure*}

In this section, we aim to assess the validity of the analytical model by comparing it to simulated data.  Here, we focus on the source galaxy bin with redshift $z=0.5$; however, in \cref{subsec:forecastsPCA} we consider also redshifts $z=1.0$ and 2.0. Although, such validation of the modeling has been addressed partly in \cite{Arvizu:2024rlt}, our focus here is to identify the scales at which the theory given by \texttt{3pt-WL} breaks down, and exclude them from the analysis to construct the data vector that is used in the following sections. 

In \cref{fig:zetam_bins}, we present the full set of bins spanned by the simulated data and for a single multipole, represented as a matrix.  We have removed the lower triangular submatrix since the multipoles are symmetric $\zeta_m(\theta_1,\theta_2)=\zeta_m(\theta_2,\theta_1)$. The bins shown in cyan are arranged into a single vector, indexed from left to right and top to bottom. We then construct a single data vector, by concatenating all computed multipoles, ranging from $m = 0$ to 8. 

In the top panel of \cref{fig:datavector_all} we show the model predictions alongside the simulated data for multipoles $\zeta_{m=0,1,2,3}$. The blue curve shows the relative difference between the mean data of the 108 simulated data and the modeling, while the cyan shading indicates the 1$\sigma$ region given by the standard deviations of the data. For small angles, corresponding to the first bins of each multipole, the model is not very accurate. In the figure, the change of multipole can be spotted as an abrupt discontinuity each 105 bins. A similar effect is seen for bins near the diagonal of the matrix of \cref{fig:zetam_bins},  corresponding to peaks in both the data and model vectors. These bins represent nearly isosceles triangles with sides $\theta_1 \approx \theta_2$, where the third side, $\theta_3$, can become arbitrarily small, and thus, such configurations are out of the reach of our analytical modeling.

In the bottom panel of \cref{fig:datavector_all}, we show the data used in this work for multipoles $\zeta_m=0$ to $\zeta_m=5$.  The data vector $\vd$, shown by the dashed black curve, is computed by averaging the signal obtained from the 108 realizations of the simulations, while the shading shows the 1$\sigma$ region around it. The modeling counterparts, $\model$, is displayed in different colors for the different multipoles. The data vector bins are chosen such that the modeling of the multipoles $\zeta_m$ from $m=0$ to 3 fall inside the 1$\sigma$ errors of the simulated data. The bins for which this occurs are indicated with a check mark in \cref{fig:zetam_bins}. The filtered bins for our analyses are also shown in the top panel of \cref{fig:datavector_all} as black circles joined by lines. Given that the errors are computed from full-sky simulations, this construction can be considered very conservative. However, this is not the case, since we will later rescale the covariance by assuming Gaussianity, and neglecting edge effects and super-sample variance. Further, we do not model baryonic physics and intrinsic alignments. The lack of all these ingredients results in overly optimistic forecasts in the upcoming sections.   

For higher multipoles, $m=4$ and 5, the bins corresponding to the smallest values of angles $\theta$ fall outside the 1$\sigma$, as can be seen from the purple and brown curves in the bottom panel of \cref{fig:datavector_all}, but well within 2$\sigma$. 
However, the error bars for this multipoles grow considerably, and, as we will see, this translate in to poor constrains of the cosmological parameters. 
In the cases of $m=6$, 7 and 8, the modeling falls within the 1$\sigma$ again; however, this is because the error bars becomes quite large compared to the amplitude of the signal. As expected, these multipoles have even smaller constraining power on cosmology.

To model the 2PCF, $\xi(\theta)$, we use the Core Cosmology Library (\texttt{CCL}) \cite{LSSTDarkEnergyScience:2018yem},\footnote{\url{https://github.com/LSSTDESC/CCL}} a publicly available standardized collection of routines for calculating fundamental cosmological observables, specialized to  theoretical predictions for the LSST Dark Energy Science Collaboration (DESC). The theoretical predictions for $\xi$, together with the data vector and 1$\sigma$ errors are shown in the upper left region of the bottom panel of \cref{fig:datavector_all}, where we have rescale them by a factor $1/2000$ to reduce the dynamical range of the figure. Unsurprisingly, the modeling is more accurate than for the $\zeta_m$'s, so we can reach even smaller scales. Hence, the $\xi$ piece of the data vector consists of 18 bins over the range $\theta=[11,200]$ arcmin spaced in logarithmic intervals. 

The final, ``complete''  data vector $\vd$ and modeling $\model$ are composed of 18 bins for the 2PCF and 45 bins for each of the multipoles considered. In total, we consider the first 9 multipoles, so the data vector consists of $N_\text{bins}=423$ bins. However, in the following analyses, we only use selected subranges of these bins and never the entire vector.

\subsection{Covariance and correlation matrices}

\begin{figure*}
	\begin{center}
	\includegraphics[width=3 in]{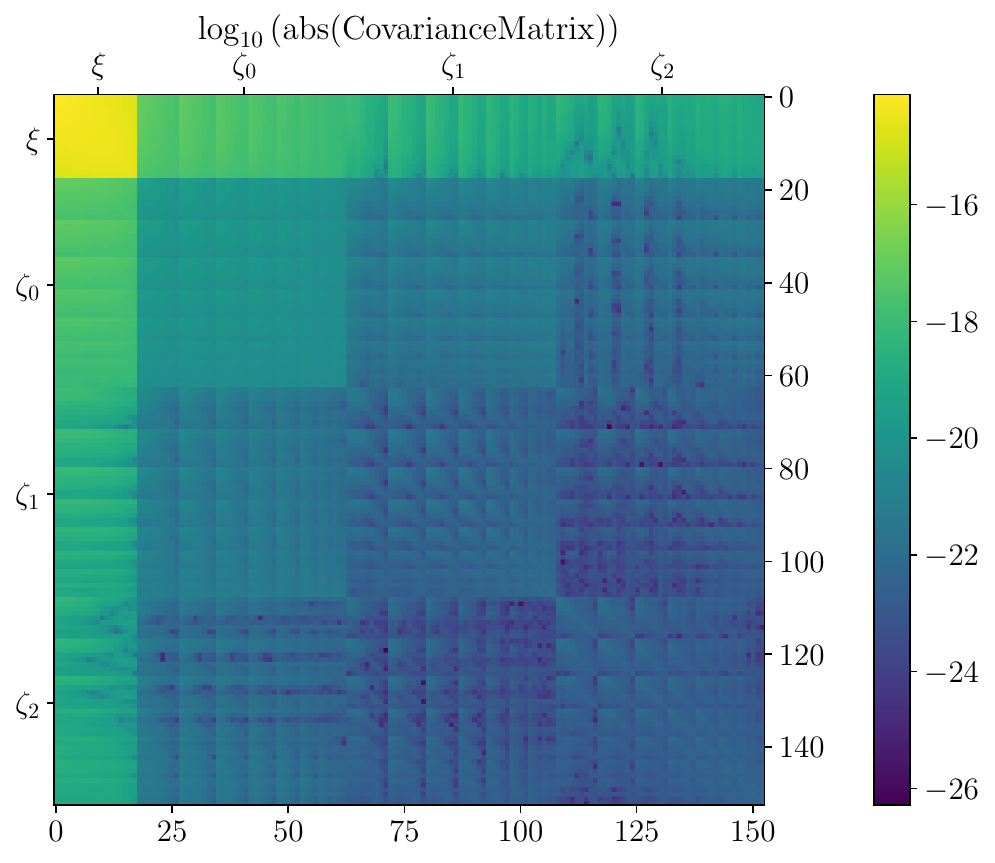}
    \includegraphics[width=3 in]{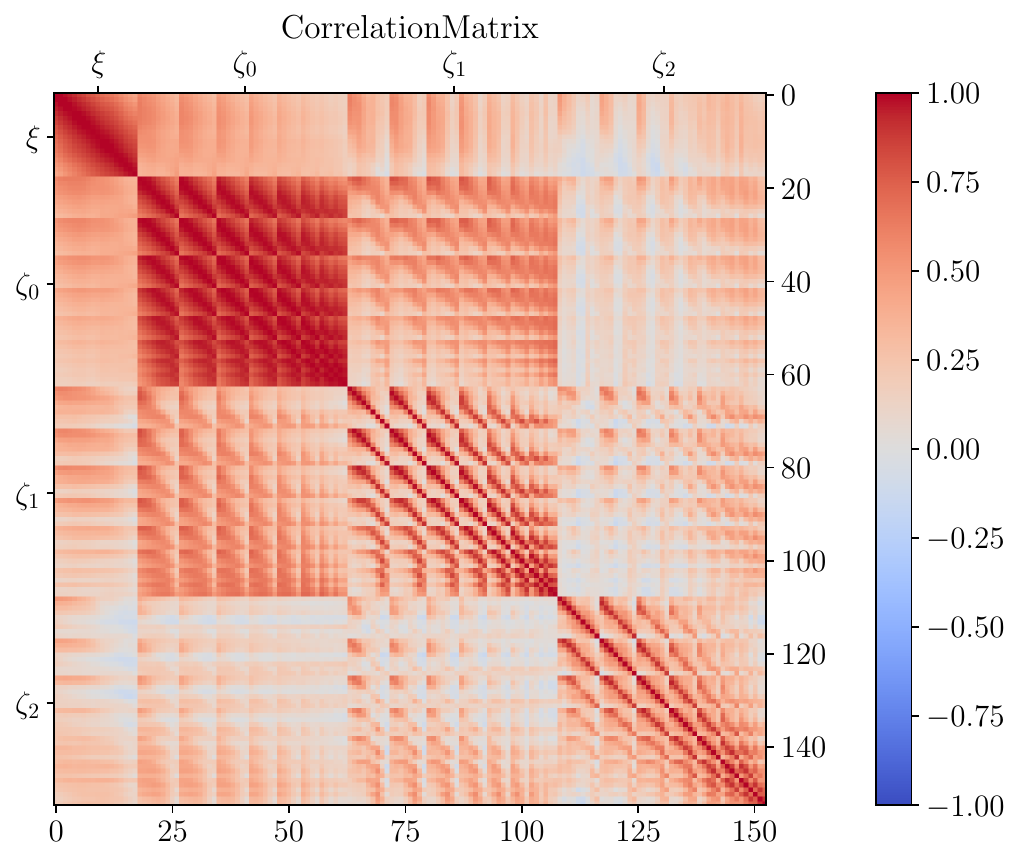}
 \caption{Sample covariance matrix (left panel) and correlation matrix (right panel) for the data vector shown in \cref{fig:datavector_all} up to bin index = 153, corresponding to the 2PCF, $\xi$, and 3PCF multipoles $\zeta_{m=0,1,2}$. Although throughout this work we use multipoles up to $m=8$, we only show $m=0,1,2$ to avoid cluttering.} \label{fig:cov_corr}
	\end{center}
\end{figure*}

We estimate the covariance matrix through the the sample covariance matrix, which we construct from the data vectors $\{\vd^{(k)}\}$, extracted from the $N_\text{sims}=108$ realizations of the full-sky lensing mock catalogs, as
\begin{equation} \label{eq:sample_cov}
    \mathbf{C} = \frac{1}{N_\text{sims}-1}\sum_{k=1}^{N_\text{sims}} (\vd^{(k)} - \vd)^T(\vd^{(k)} - \vd),
\end{equation}
where  $\vd=N_\text{sims}^{-1} \sum_{k=1}^{N_\text{sims}} \vd^{(k)} $ is the sample mean. We note that the number of bins in the data vector is larger than the number of simulations, and hence $\mathbf{C}$ is not a  \textit{valid} covariance matrix; in particular, it is singular \cite{Hartlap:2006kj}. However, we will never use $\mathbf{C}$ in its full form. Instead, we will apply reductions to it, that will allow us to use it for forecasting.

The left panel of \cref{fig:cov_corr} shows the covariance matrix for the data vector presented in the bottom panel of \cref{fig:datavector_all}. We plot the logarithm of the covariance because its values span over 13 orders of magnitude when including both $\xi$ and $\zeta_m$; for comparison, the 3PCF alone spans over 10 orders of magnitude. To avoid cluttering, the plot only shows the covariance including $\xi$ and multipoles up to $m=2$, leaving aside higher multipoles $m=3,\cdots \!,8$. Despite this and the use of a logarithmic scaling, the structure of the submatrices corresponding to the auto-covariance of the individual components is not clearly revealed in the figure, as is more evident for the  $\xi\!-\!\xi$ block. However, it can be noticed that this submatrix is isolated from the rest of the covariance, a feature that will be relevant when we perform the PCA in \cref{sec:PCA}.  
 We also notice that the cross-covariance values between different multipoles are considerably large, certainly larger than the cross-covariance between multipoles and the 2PCF.  

In \cref{sec:GaussianCov}, we will compare these results with an analytically derived covariance under the assumption of Gaussianity, finding overall reasonable agreement but also some notable differences.

The right panel of \cref{fig:cov_corr} displays the correlation matrix.  The high values in the cross-terms indicate that correlations are significant between different bins, not only within the same multipole, but also across them. This shows a high redundancy among different multipoles, an important factor that will lead us to use a PCA to study the multipoles altogether in \cref{sec:PCA}. We also note that while the auto-correlations are predominantly positive for the lowest multipoles, beginning at $m=4$ (not shown in the plot) the positive and negative correlations become more evenly distributed away from the main and adjacent diagonals.

\section{Cosmological forecasts per multipole}\label{sec:Forecast1pole}

\begin{figure*}
	\begin{center}
    \includegraphics[width=6 in]{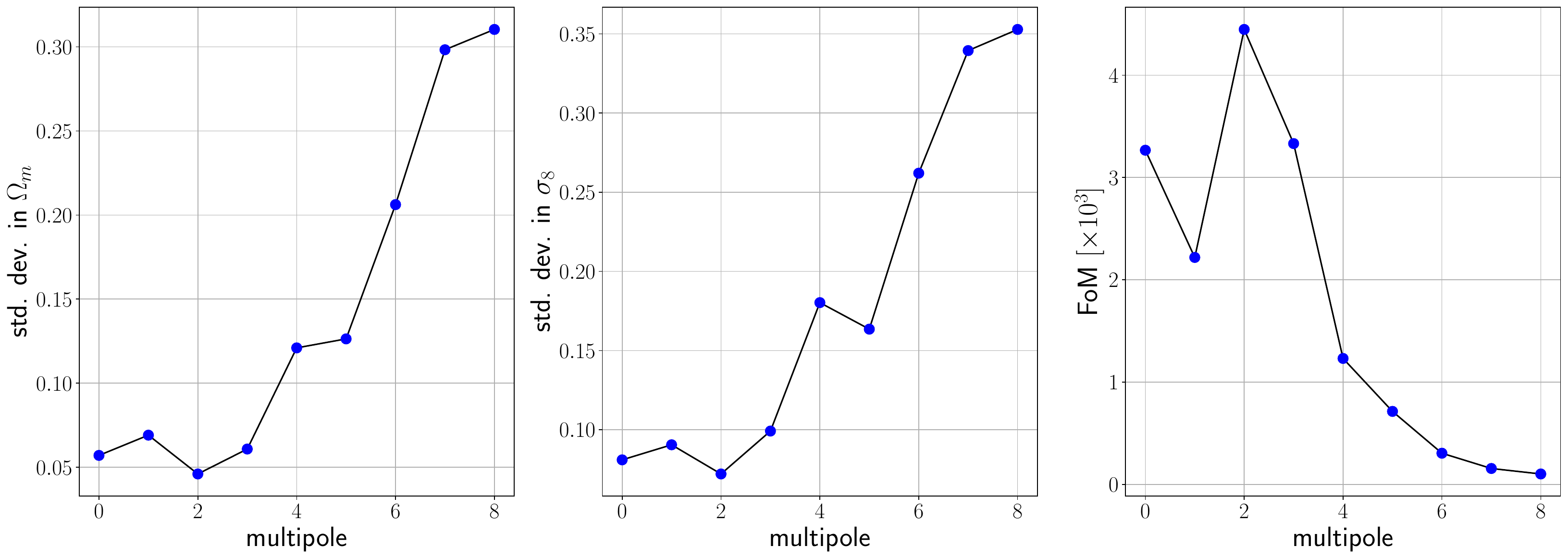}
 \caption{Fisher forecast over the $(\Omega_m, \sigma_8)$ parameter space when the multipoles from $\zeta_{m=0}$ to $\zeta_{m=8}$ are analyzed separately. The left and center panels show the $1\sigma$ errors on $\Omega_m$ and $\sigma_8$, respectively, while the right panel shows the FoM. This plot shows the quadrupole is the multipole providing the most restrictive constraints, with $\sqrt{{\rm FoM}}=66.69$. For comparison, the value for square root of the FoM for the 2PCF is 118.71.} \label{fig:errors_1}
	\end{center}
\end{figure*}

In this section we present forecasts for different multipoles alone in the space $(\Omega_m, \sigma_8)$.\footnote{As is standard, in practice we vary $A_s$ and rescale $\sigma_8$ with the rest of cosmological parameters fixed.} Having a log-likelihood function $\mathcal{L}(\vd| \vp)$ for cosmological parameters $\vp=\{ p^1,p^2,\dots \}$ and data vector $\vd$, the Fisher information matrix around $\bar{\vp}$ is given by 
\begin{equation}
    F_{ij} =- \left\langle \frac{\partial^2  \mathcal{L}}{\partial p^i \partial p^j} \right\rangle \Bigg|_{\vp=\bar{\vp}}. 
\end{equation}
For a Gaussian likelihood, and assuming the covariance does not depend on the paramters $\vp$, the Fisher matrix becomes 
\begin{equation}
    F_{ij} = \frac{\partial \mu^a}{\partial p^i} \frac{\partial \mu^b}{\partial p^j} C^{-1}_{a b},
\end{equation}
where sum over repeated indices is assumed. The indices $a$ and $b$ run over all bins on the data vector $\vd$, while $i$ and $j$ over the cosmological parameters. The covariance matrix $C$ is obtained from the simulations as outlined in the previous section, while the model vector and its derivatives is obtained analytically using the codes \texttt{3pt-WL} and $\texttt{CCL}$.

In the left and center panels of \cref{fig:errors_1} we show the errors on $\Omega_m$ and $\sigma_8$ obtained from $1/\sqrt{F_{11}}$ and  $1/\sqrt{F_{22}}$, respectively. In the right panel we show the Figure of Merit (FoM) given by $\text{FoM}=\det(F_{ij})$. We notice that the quadrupole is the most constraining multipole. This is a consequence of the large signal presented in the quadrupole and the small dispersion of the data around their mean, as observed in \cref{fig:datavector_all}. The second most constraining multipole is the monopole. For $m>2$, the FoM falls to zero rapidly, being the overall error ratio $\big(\text{FoM}(m=8)/\text{FoM}(m=2)\big)^{1/2} \sim 6$.  The Fisher forecast contours at 1 and 2$\sigma$ confidence level for the quadrupole are shown with black-dashed curves in the left panel of \cref{fig:s8toS8}.

\subsection{Beyond Fisher and $S_8$}
\label{subsec:beyondF}

\begin{figure*}
	\begin{center}
    \includegraphics[width=3 in]{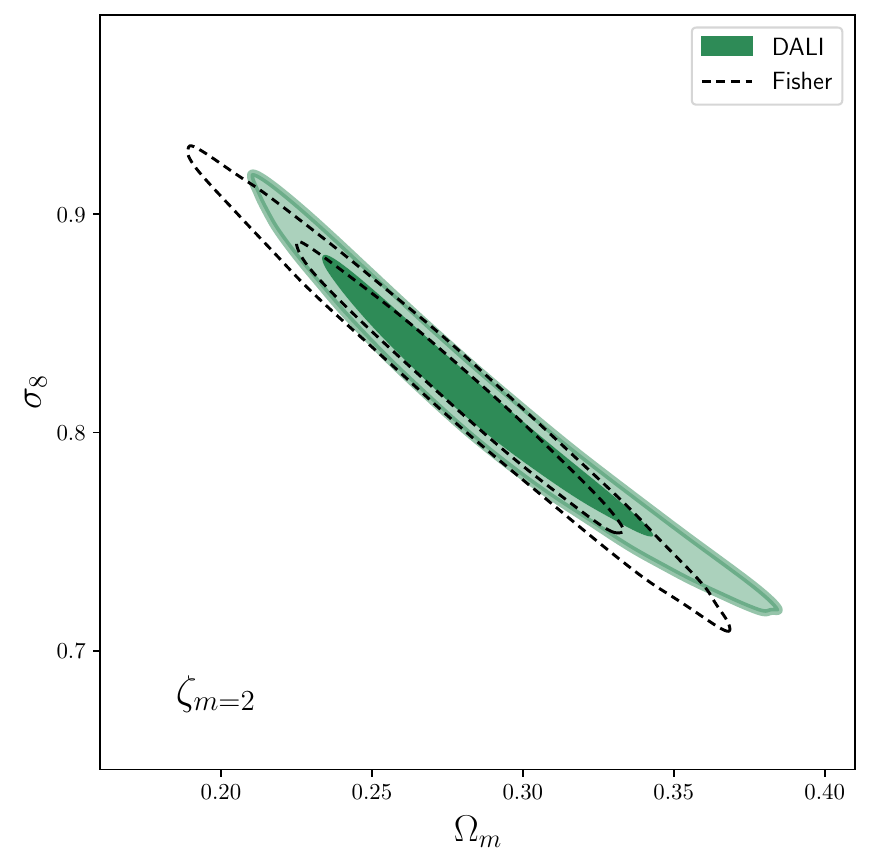}    \includegraphics[width=3 in]{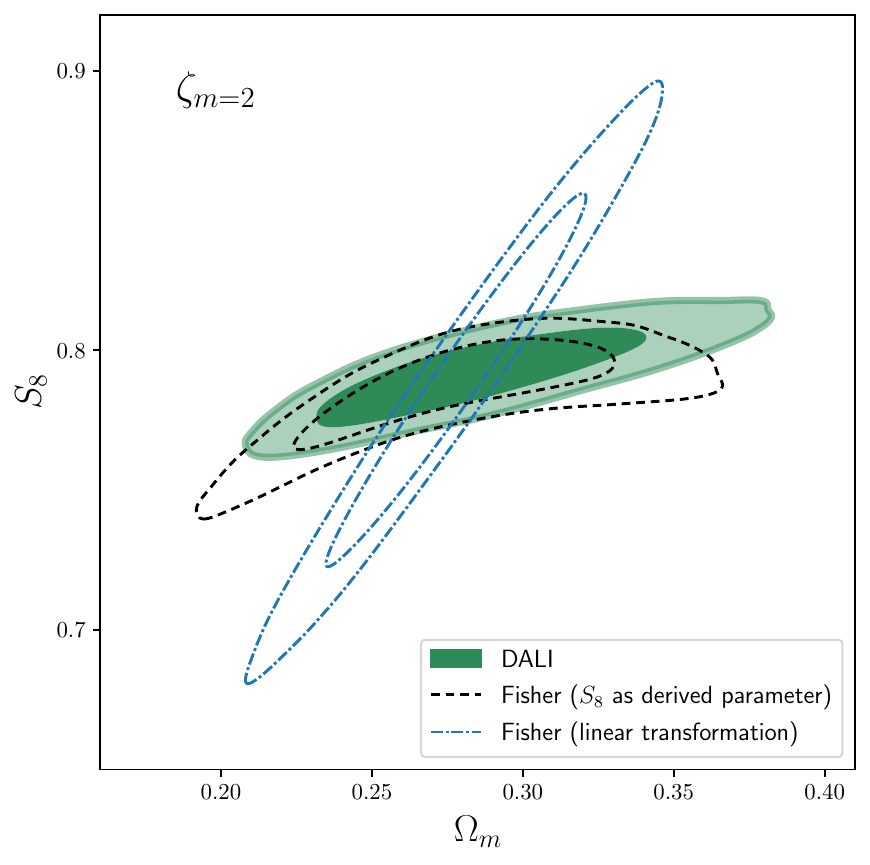}
 \caption{\textit{Left panel:} 1 and 2 $\sigma$ confidence contours on parameter space $(\Omega_m,\sigma_8)$, provided by forecasting with Fisher (black-dashed curves) and DALI (green contours). \textit{Right panel:} forecasts on the space of parameters $(\Omega_m,S_8)$, where $S_8$ is computed from the chains obtained using DALI (green contours) and using Fisher (black dashed curves). We also show in the dot-dashed blue curves the contour obtained by transforming the Fisher matrix with the linear relation in \cref{FisherLT}. In both panels we have used the quadrupole of the 3PCF for the galaxy distribution centered at redshift $z=0.5$.} \label{fig:s8toS8}
	\end{center}
\end{figure*}

Weak lensing is primarily sensitive to the amplitude of matter fluctuations projected along the line of sight, which depends on both the growth of structure and the geometry of the universe. This leads to a strong degeneracy between $\Omega_m$ and $\sigma_8$ in the lensing signal. The parameter $S_8 = \sigma_8 \sqrt{\Omega_m/0.3}$ captures this degeneracy direction, providing a more direct and tighter constraint from lensing data. 
Further, since the degeneracy in the  $(\Omega_m,\sigma_8)$ space is non-linear, a Fisher analysis cannot capture its characteristic ``flexed'' shape. Hence, when transformed to  $(\Omega_m,S_8)$ space the outcomes become unreliable. This is shown with the dashed black curves in \cref{fig:s8toS8}. In the left panel we have implemented the Fisher analysis, for which we have ran Monte Carlo Markov Chains (MCMC) drawing the probability distribution\footnote{To do this we use the code \texttt{emcee} \cite{emcee}, available at \url{https://github.com/dfm/emcee}. The chains are analyzed with the \texttt{GetDist} package \cite{Lewis:2019xzd}, available at \url{https://github.com/cmbant/getdist}. }
\begin{equation}\label{prob_F}
    \mathcal{P}_F(\vp) = \exp\left(-\frac{1}{2} \Delta_i\Delta_j F_{ij} \right),
\end{equation}
where $\vp=(\Omega_m,\sigma_8)$ and $\Delta= \vp - \bar{\vp}$ with the fiducial parameters of the simulations $\bar{\vp}=(0.279,0.82)$. We adopt the approach of sampling the probability distribution given by \cref{prob_F}, instead of using the Fisher matrix directly to plot the forecasted contours, because it allows us to easily compute $S_8$ as a derived parameter and to generalize the method beyond the linear approximation of Fisher analysis. 
%Having the chains at hand, we can easily compute $S_8$ as a derived parameter. 
In the right panel of \cref{fig:s8toS8}, we show its distribution (dashed black curve), which exhibits a highly non-linear shape. To generate \cref{fig:s8toS8}, we use the quadrupole component of the vector data, $\zeta_2$, as a representative multipole of the 3PCF, in addition that it is the most constrictive one.

To go beyond Fisher, we follow the Derivative Approximation for LIkelihoods (DALI) method \cite{Sellentin:2014zta}, see also \cite{Sellentin:2015axa,Ryan:2022qpa}, that expands a log-probability distribution $\log(\mathcal{P})$ in derivatives of the data (or model) vector, instead than in $\vp$ itself. To implement DALI, we compute the second derivatives of the model vector with respect to the cosmological parameters and multiply them by the inverse of the covariance matrix as
\begin{align}
    G_{ijk} &= \frac{\partial^2 \zeta^a}{\partial p^i \partial p^j} \frac{\partial \zeta^b}{\partial p^k} C^{-1}_{ab},  \\
    H_{ijkl} &= \frac{\partial^2 \zeta^a}{\partial p^i \partial p^j} \frac{\partial^2 \zeta^b}{\partial p^k \partial p^l} C^{-1}_{ab}. 
\end{align}
We then construct the probability distribution
\begin{equation}\label{prob_2}
    \mathcal{P}_D(\vp) = \exp\left(-\frac{1}{2} \Delta_i\Delta_j F_{ij} -\frac{1}{2} \Delta_i\Delta_j \Delta_k G_{ijk} -\frac{1}{8} \Delta_i\Delta_j \Delta_k\Delta_l H_{ijkl} \right),
\end{equation}
which we sample using the MCMC hammer  method \cite{emcee}. The resulting filled contour, shown in green in the left panel of \cref{fig:s8toS8}, exhibits the characteristic shape typically inferred from weak lensing data. Next, we map the distribution to the $(\Omega_m, S_8)$ space by directly manipulating the chains, both for DALI (green contour) and Fisher (dashed black lines). We further transform parameters using the linear relation  
\begin{equation} \label{FisherLT}
 \tilde{F}_{ab} =  \frac{\partial \tilde{p}^a}{\partial p^i} \frac{\partial \tilde{p}^b}{\partial p^j} F_{ij},  
\end{equation}
where the tilde parameters are $ \tilde{\vp}=(\Omega_m,S_8)$. The resulting contours are shown with blue dot-dashed curves.  

% $\tilde{F}_{ab} = \partial_{}$, 
% the degeneracy becomes linear. 
%
Despite the exponent in \cref{prob_2} being positive definite, when we explore a large space of parameters, beyond  $(\Omega_m, \sigma_8)$, we have encountered numerical difficulties that brought the term $G=-\frac{1}{2} \Delta_i\Delta_j \Delta_k G_{ijk}$ sufficiently large and positive, such that the probability $\mathcal{P}_D$ becomes larger than unity. Besides that, we have  
obtaining spurious bimodal distributions.\footnote{In the one-dimensional case, the exponent in \cref{prob_2} becomes $-\frac{1}{2} (\Delta d')^2 -\frac{1}{2}\Delta^3 d' d'' -\frac{1}{8}(\Delta^2 d'')^2 $, up to a positive constant. This form always leads to bimodalities in $\mathcal{P}_D$, because the polynomial has two maxima as long as $d'$ and $d''$ are different from zero.} This and others limitations of the DALI method are highlighted in \cite{Ryan:2022qpa}. %However we will comeback to it in the next section to show the contours our "final" data vector in Sec.XXX. 
%
%\url{https://arxiv.org/pdf/1205.3984}
%
%\url{https://arxiv.org/pdf/2211.06534}
%
In contrast, within the Fisher approximation we never find bimodalities,  because $F^{-1}$ is the matrix of a positive-definite quadratic form. In this sense, while going beyond Fisher provides more accurate approximations to the true distributions, it requires additional care. However, the Fisher approximation itself is expected to break down for large departures from the mean values of the cosmological parameters \cite{Wolz:2012sr}, even before the DALI approximation fails.

\section{Principal component analysis}\label{sec:PCA}

\begin{figure*}
	\begin{center}
    \includegraphics[width=6 in]{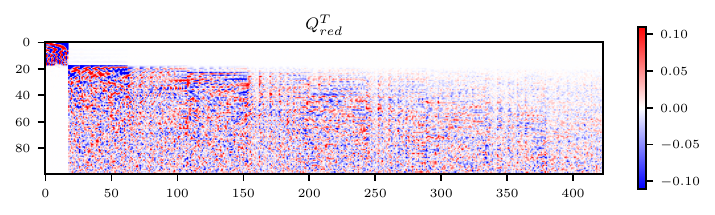}
 \caption{$Q$ matrix that transform vectors from the original to the PCA space.  The upper-left block corresponds to the 2PCF, while the lower-right block corresponds to the 3PCF. The cross terms are very small, indicating that the mixing between $\xi$ and $\zeta$ components is almost negligible.  } \label{fig:Q_red}
	\end{center}
\end{figure*}

\begin{figure*}
	\begin{center}
    \includegraphics[width=6 in]{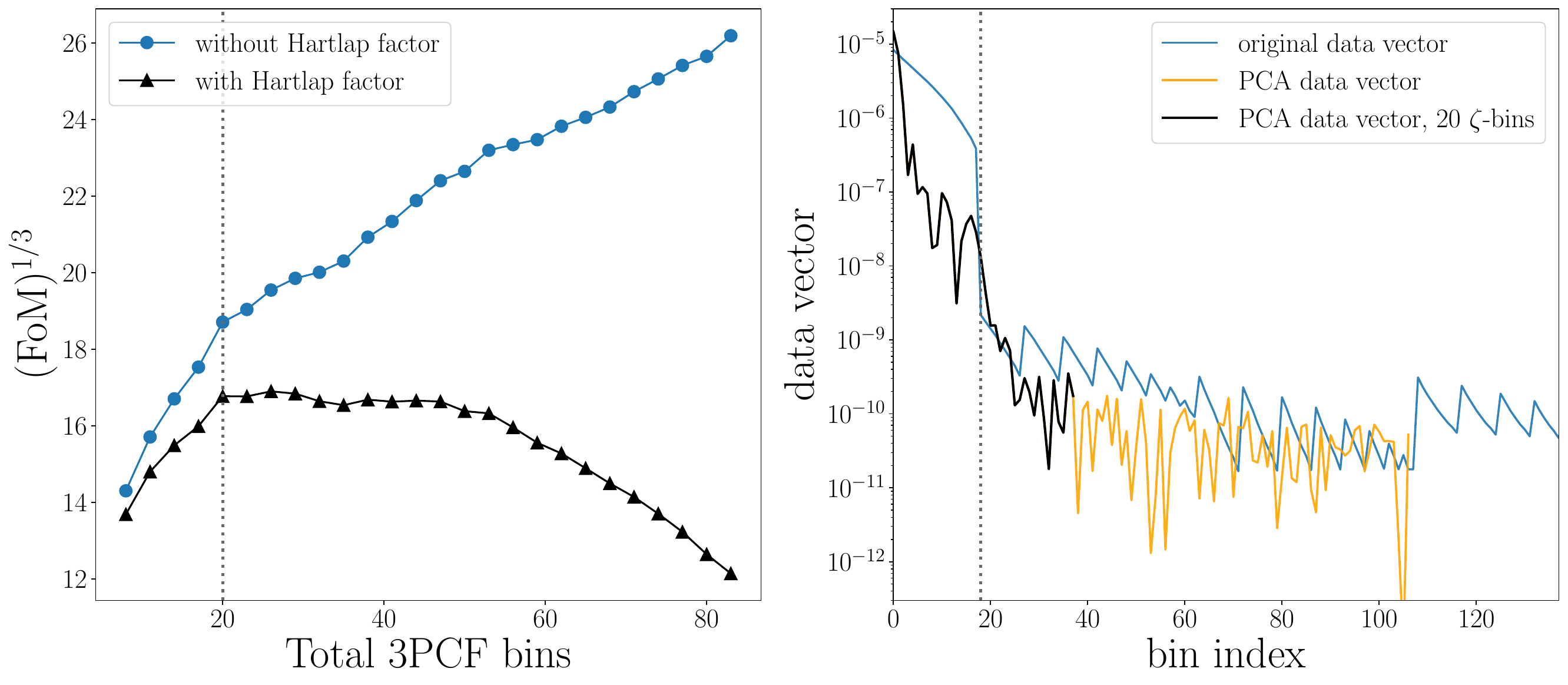}
 \caption{Principal Component Analysis compression. \textit{Left panel:} cube root of the FoM obtained from the Fisher forecast as a function of the number of components kept in the 3PCF multipoles. We show results with (triangles) and without (circles) considering the Hartlap factor. The dashed vertical line show the value shown in the rest of this work. \textit{Right panel:} The PCA data vector including the 20 3PCF components and the 18 2PCF is shown with the black curve. The orange curve shows the PCA data vector with maximal variance that can be obtained from the 108 simulations. The blue curve shows the original data vector up to the bin index 140. The vertical dashed line separates the data vectors into the 2PCF portion (left hand side) and the 3PCF portion (right hand side).
% El vector dPCA 20 bins tiene un el 0.996232312275144 de la varianza total 
 } \label{fig:FisherForecast}
	\end{center}
\end{figure*}

A challenge in our analysis is to estimate a reliable inverse covariance matrix that allows us to perform forecasts using high-dimensional data vectors. In our case, the full data vector consists of $N_\text{bins}=423$ elements, obtained from measurements of 18 bins from the 2PCF and 45 bins from each of the first 9 multipoles of the 3PCF. However, we only have access to 108 independent full-sky simulations to estimate the covariance. It is known from \cite{Pan:2005ym,Hartlap:2006kj} that the sample covariance matrix becomes singular when the number of simulations $N_{\text{sim}}$ is smaller than the number of data bins $N_\text{bins}$. Even when $N_\text{sim}$ slightly exceeds $N_\text{bins}$, the inverse covariance is still noisy and biased. 
In our case, the covariance $\mathbf{C}$ introduced in  \cref{eq:sample_cov} is strictly singular and cannot be used for parameter inference or Fisher forecasting in its full form.
Moreover, the correlation matrix shown in \cref{fig:cov_corr} reflects that the data vector contains a high degree of redundancy. This further reduces the usefulness of treating all 423 bins and motivates a more compact representation. To address both the mathematical limitations and the informational redundancy, we apply PCA \cite{AbdiWilliams2010} to reduce the dimensionality of the data vector.  By using this procedure the compressed data vector is composed of statistical independent modes and suppresses directions dominated by noise or degeneracies.

Now, our purpose is to test how many components can we retain in the PCA of the 3PCF without losing cosmological information. Henceforth, for the moment we do not consider the 2PCF bins in the data vector, and decompose the covariance matrix as $\mathbf{C} = \mathbf{Q} \mathbf{\Lambda}  \mathbf{Q}^T$, where $\mathbf{\Lambda}$ is the $N_\text{bins}\times N_\text{bins}$ diagonal matrix of eigenvalues sorted in descending order $\lambda_1 \geq \lambda_2 \geq \cdots \geq \lambda_{N_\text{bins}}$, and $\mathbf{Q}$ is the matrix of eigenvectors accommodated in columns. In our case, the last $N_\text{bins}+ 1 -N_\text{sims}$ are all zero, so we are limited to a maximum of 107 components. 

We construct the ``reduced'' matrix $\mathbf{Q}_\text{red}$ by removing from $\mathbf{Q}$ the last $N_\text{bins}-N_\text{sims}$ rows.  In \cref{fig:Q_red} we show its transpose which becomes an $n \times N_\text{sims}$ matrix used to transform each data vector from the simulations, $\vd^{(k)}$, and the model, $\mu(p)$, into compressed, or PCA, $n$-dimensional versions:
\begin{equation} \label{vtoPCA}
\vd_{\rm PCA} = \mathbf{Q}_{\text{red}}^T \vd. % \quad \vd_{\rm PCA}^{{\rm sim},i}  = \mathbf{Q}_{\text{red}}^T \vd_{\rm PCA}^{{\rm sim},i} .
\end{equation}

In \cref{fig:Q_red} we consider $n$ equal to 107. However we can reduce the number of entries on the $\vd_{\rm PCA}$ data vectors, by removing more eigenvectors from $\mathbf{Q}_\text{red}$. That is, by removing  principal components from the 3PCF multipoles corresponding to the smaller eigenvalues. In the left panel of \cref{fig:FisherForecast}, we plot the FoM computed with the Fisher analysis by changing the number of PCA components kept in the 3PCF. To this end, we consider the space of parameters $(\Omega_m,\sigma_8,w_0)$. The blue circles show the results when considering the covariance directly from  \cref{eq:sample_cov}, while for plotting the black triangles and lines we include the Hartlap factor to the inverse covariance, by rescaling \cite{Hartlap:2006kj}
\begin{equation}
    C^{-1} \rightarrow \frac{N_\text{sims} - N_\text{bins} - 2}{N_\text{sims}-1} C^{-1}.
\end{equation}
The Hartlap correction is necessary to obtain unbiased cosmological forecasts, and it has a particular relevance when the number of data bins approaches the number of simulations used to estimate the covariance matrix. For this reason, \cref{fig:FisherForecast} helps us to decide how many PCA components to retain from the 3PCF, finding that selecting around 20 components of the 3PCF yields the highest FoM, hence the greatest constraining power.  Choosing 40 components yields similar results; however, the Percival factor penalization to the final constraints \cite{Taylor:2012kz,Dodelson:2013uaa,Percival:2013sga}, due to the propagation of errors in the covariance matrix to the cosmological parameters, %which is not accounted for in \cref{fig:FisherForecast}, 
is larger in such case. %, since it goes as the difference between the number of simulations and bins.

We then construct the final PCA compressed data vector by including the 18 bins from the 2PCF and computing the transformation matrix $\mathbf{Q}_\text{PCA}$ using the first 38 eigenvectors of the sample covariance matrix. In the right panel of \cref{fig:FisherForecast}, the blue curve shows the first 200 entries of the full 423 dimensional data vector $\vd$, the orange line the 107 dimensional compressed vector, which corresponds to the maximum dimensionality allowed by our 108 simulations,  and our final 38 dimensional PCA data vector, $\vd_\text{PCA}$, is shown with the  black curve.  

As shown in \cref{fig:Q_red}, the matrix $\mathbf{Q}_{\text{red}}$ exhibits a clear block-diagonal structure with near-zero values outside the blocks. The upper-left block corresponds to 2PCF bins, while the bottom-right block corresponds to 3PCF bins. This structure shows a strong mixing within 2PCF and 3PCF components but negligible mixing between them. Hence, in the right panel of \cref{fig:FisherForecast}, the 18 PCA components shown to the left of the dashed vertical line correspond mainly to the 2PCF, while the 20 on the right belong to the 3PCF, giving a total of 38 bins.

In the next subsection we produce the forecasts with PCA. To do this, we will analyze different redshifts, which have slightly different settings.

\subsection{Forecasts with PCA}
\label{subsec:forecastsPCA}

\begin{figure*}
	\begin{center}
    \includegraphics[width=1.9 in]{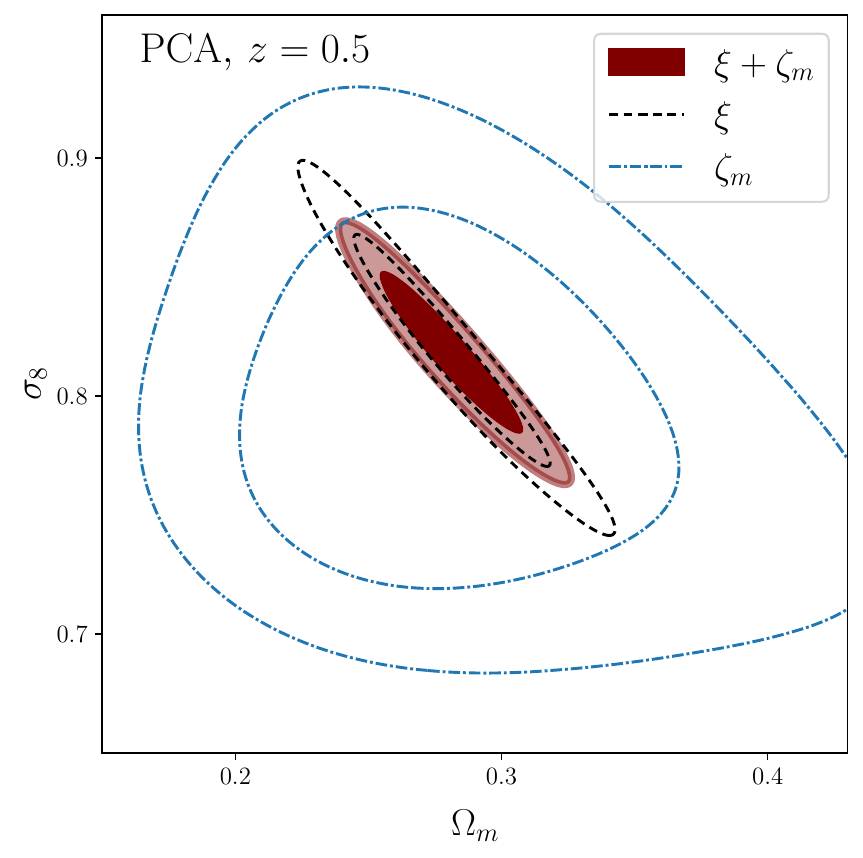}
    \includegraphics[width=1.9 in]{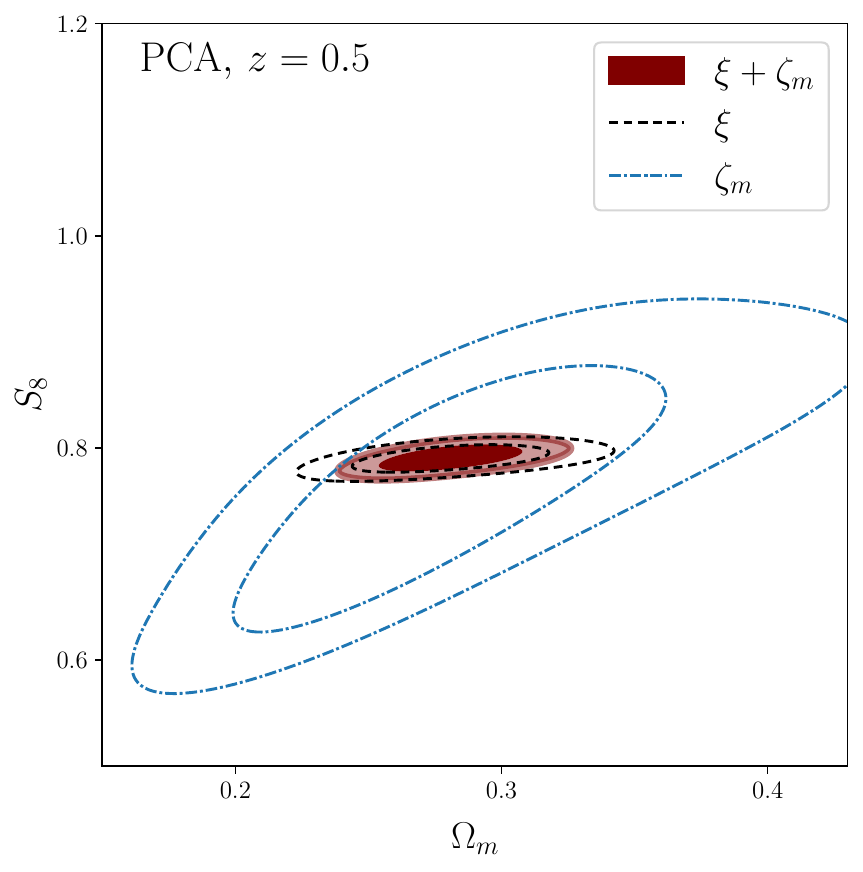}
    \includegraphics[width=1.9 in]{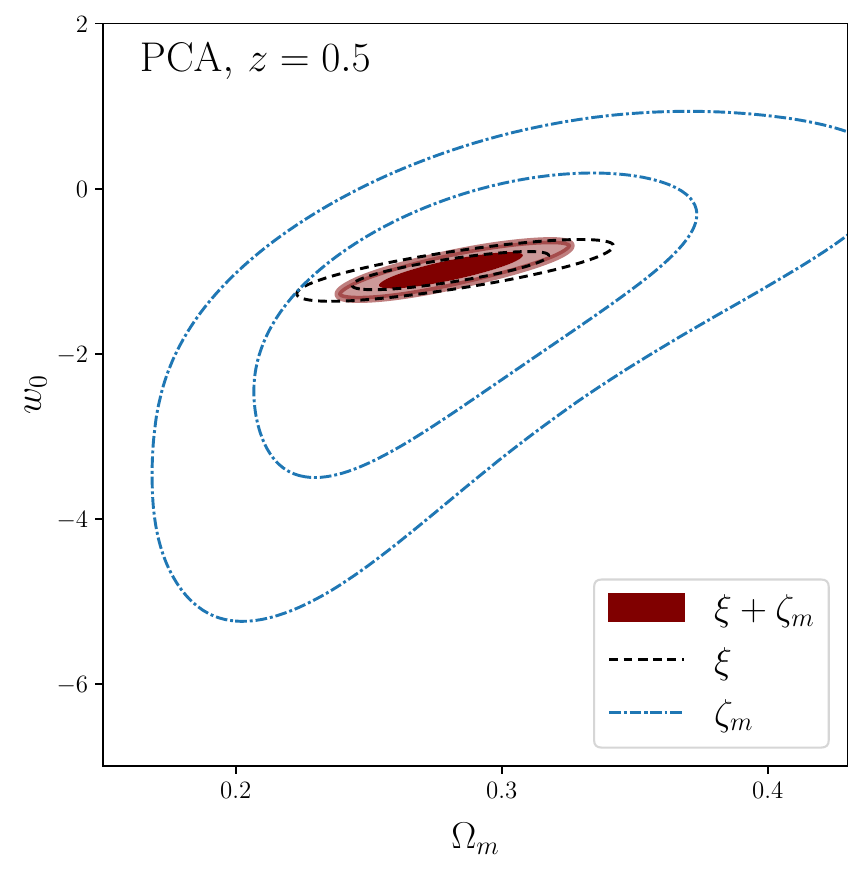}
    \includegraphics[width=1.9 in]{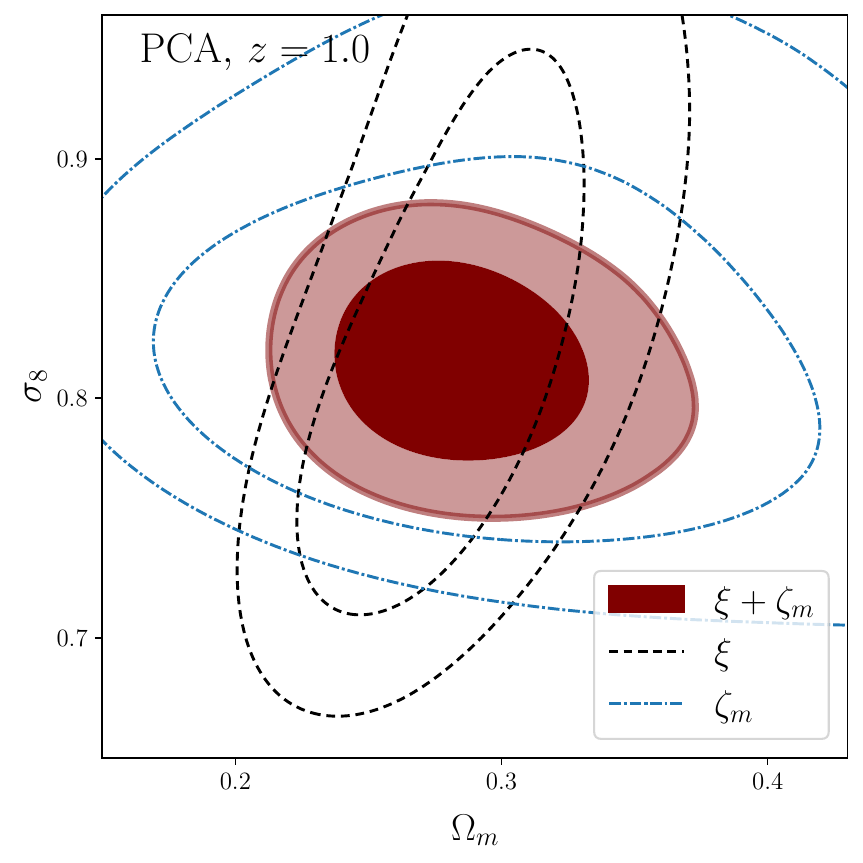}
    \includegraphics[width=1.9 in]{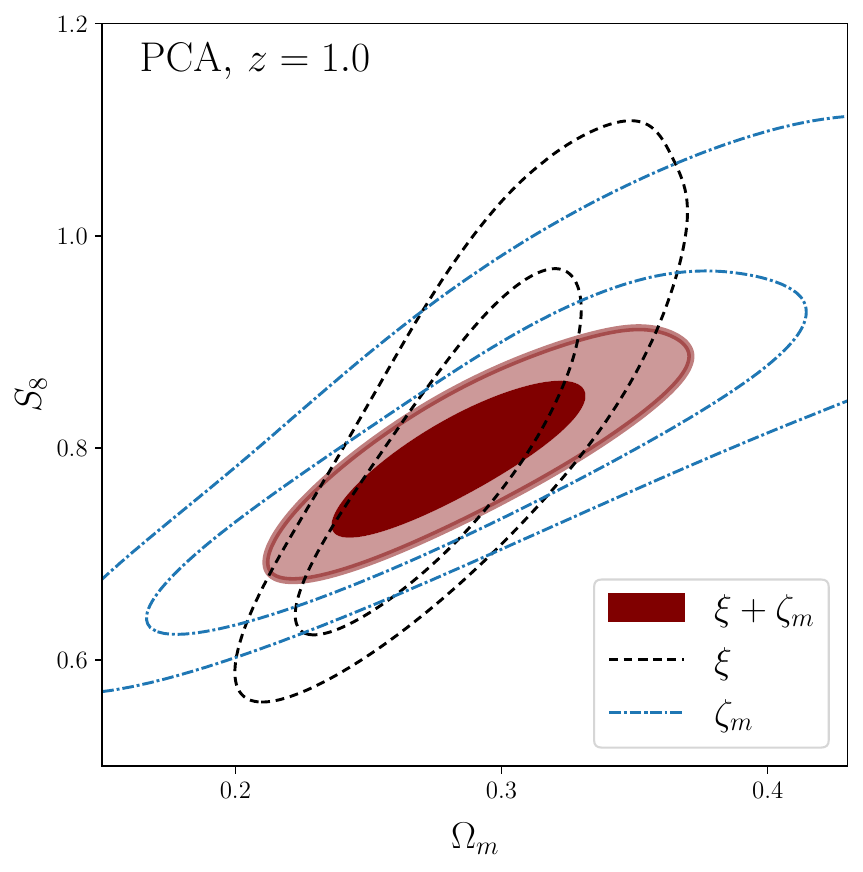}
    \includegraphics[width=1.9 in]{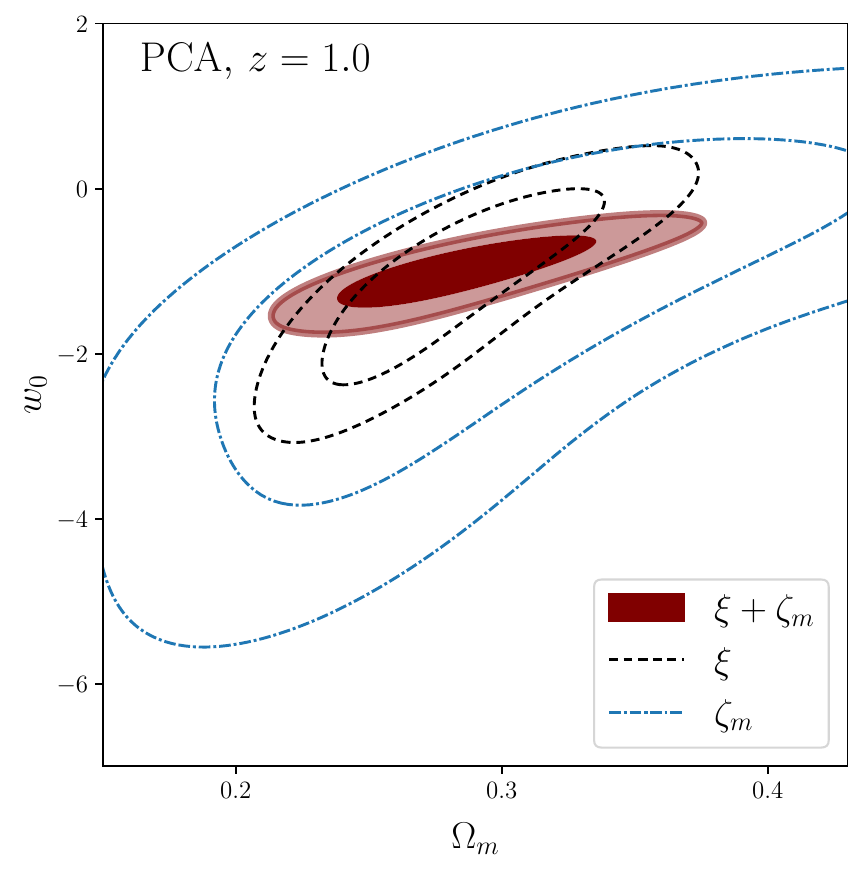}
    \includegraphics[width=1.9 in]
    {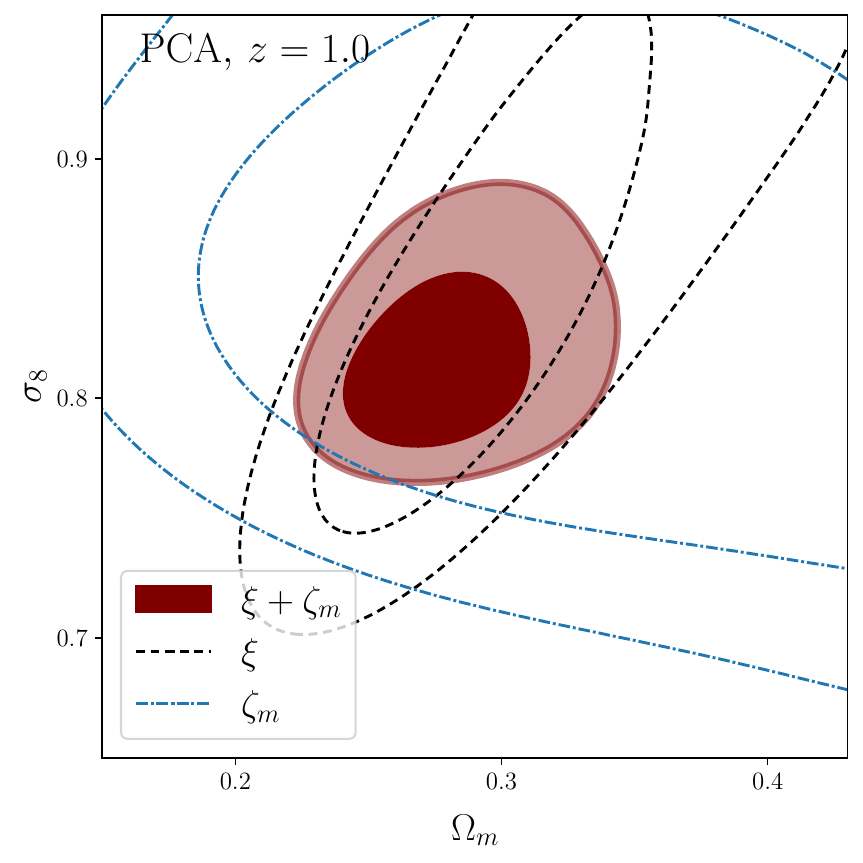}
    \includegraphics[width=1.9 in]{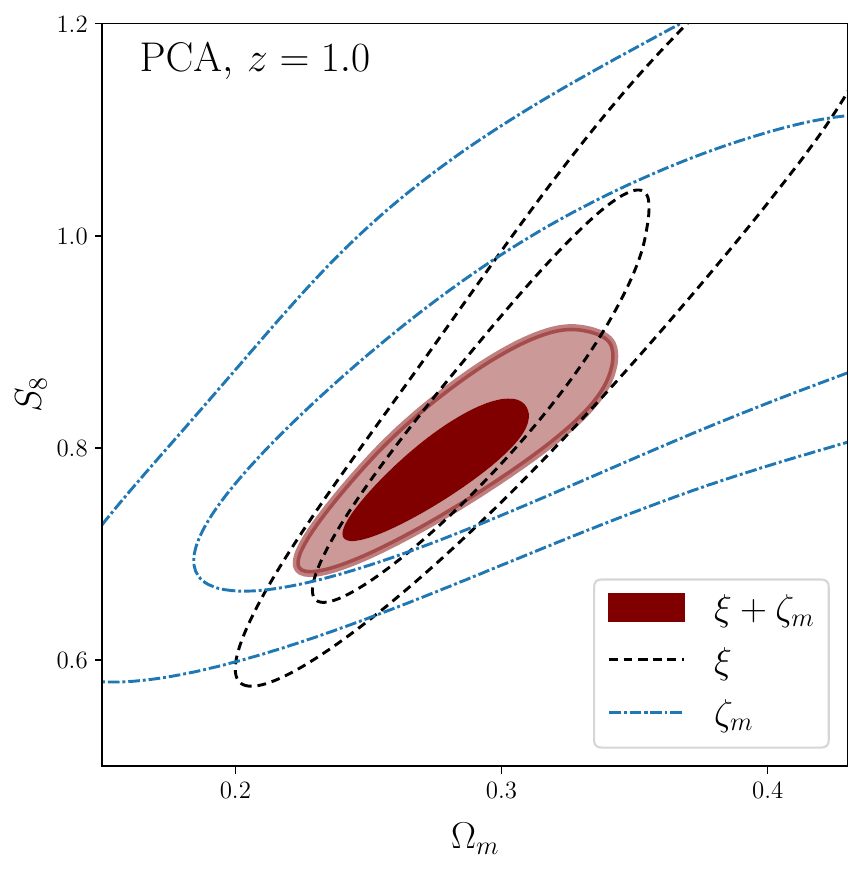}
    \includegraphics[width=1.9 in]{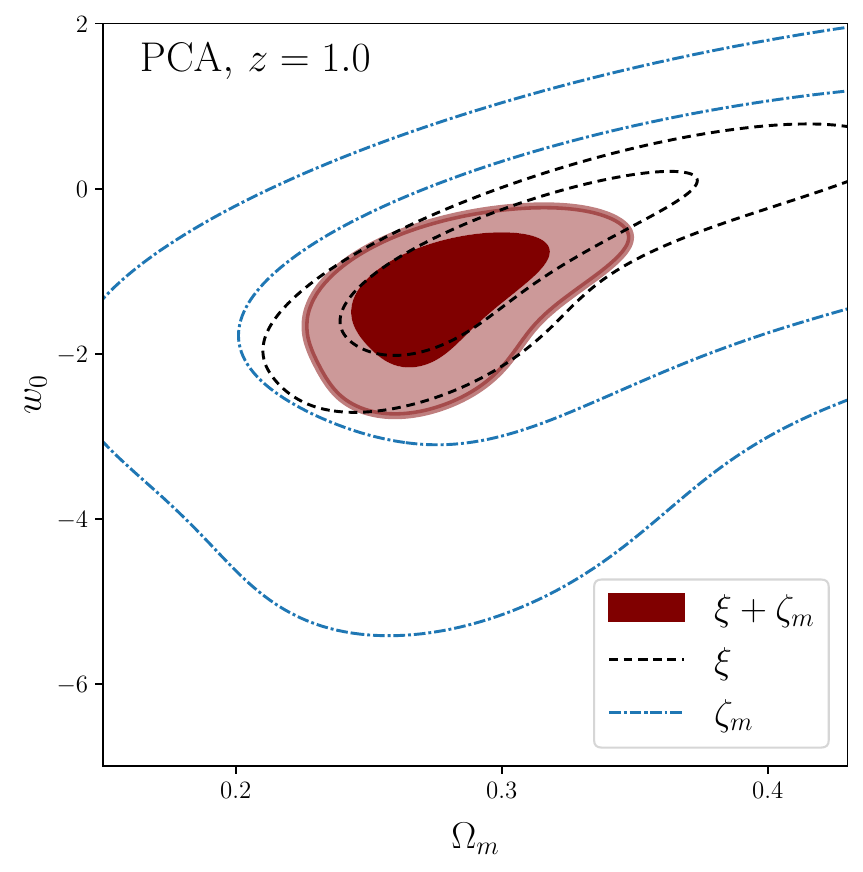}
 \caption{1 and $2\sigma$ contour plots for parameters $(\Omega_m,\sigma_8,S_8,w_0)$. We show the cases of the 2PCF alone (black-dashed curves), the 3PCF alone (dot-dashed blue curves) and the combined 2PCF+3PCF  (red contours). The improvement provided by the 3PCF in parameters $\Omega_m$ and $\sigma_8$ are around the 20\% for galaxy sources at $z=0.5$ (top panels). In the middle and bottom panels we show the samples at galaxy sources $z=1.0$ and $z=2.0$, respectively. We use the PCA compression considering 20 components from the 3PCF and 18 from the 2PCF.
 } \label{fig:tri_zs9}
	\end{center}
    
\end{figure*}

\begin{figure*}
	\begin{center}
  \includegraphics[width=5.2 in]{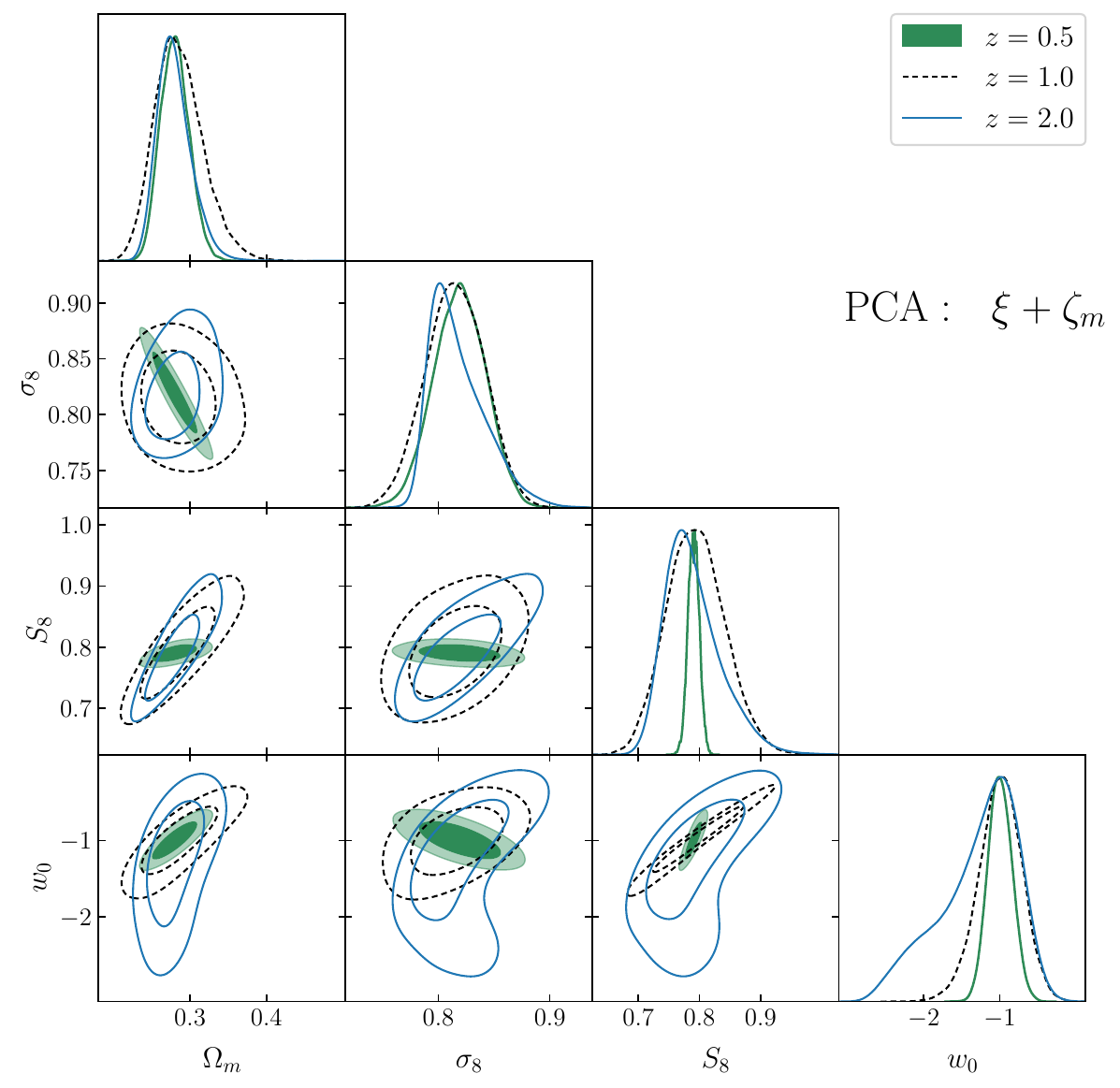}
 \caption{DALI forecasts for the PCA data vector constructed from the combined 2PCF ($\xi$) and 3PCF multipoles ($\zeta_m$). We show results for source galaxy distributions at $z=0.5$, 1.0 and 2.0.  } \label{fig:results_allz}
	\end{center}
\end{figure*}

In our final configuration, for redshift $z=0.5$, we use a compressed data vector with 38 bins, and a suite of 108 simulations to estimate the covariance matrix. This gives a simulation-to-bin ratio of $2.84$ and a Hartlap correction factor of $0.65$. While this is below the optimal threshold for high-precision inverse covariance estimation, it allows us to capture its structure and is acceptable for our primarily qualitative analysis. Our focus is on understanding the relative information content of the 3PCF with respecto to that of the 2PCF, rather than deriving precise parameter forecasts. In the upcoming section, we will complement this approach with an analytical Gaussian covariance model.

Hence, here we present DALI forecasts with parameters $(\Omega_m,\sigma_8,S_8,w_0)$ for the PCA data vector. 
To construct the probability distributions of \cref{prob_2}, we apply the transformation \cref{vtoPCA} to the model vector $\model$ and its first and second derivatives with respect to the cosmological parameters  $\vp = \{\Omega_m,\sigma_8,w_0\}$.

We rescale the covariance matrix by the ratio of the full-sky to half-sky area. This brings our analysis closer to the expected footprints of next-generation surveys like LSST ($A_{\text{LSST}} \sim 18,\!000\,\text{deg}^2$) and Euclid ($A_{\text{Euclid}} \sim 14,\!500\,\text{deg}^2$), which cover approximately 44\% and 34\% of the sky, respectively. Specifically, we apply the transformation
\begin{equation}
    C \rightarrow \frac{1}{f_{sky}} C \, , \quad \text{with} \quad f_{sky} = 0.5 \, ,
\end{equation}
increasing the statistical uncertainties by a factor of $\sqrt{2}$.

We use MCMC to sample the probability distribution $\mathcal{P}_D$ to obtain chains for the set of parameters $\vp$, and we further compute $S_8$ as a derived cosmological parameter. The resulting forecasts for source galaxies at redshift $z = 0.5$ is shown in the top panels of \cref{fig:tri_zs9}. The blue dot-dashed lines show constraints from the 3PCF multipoles $\zeta_m$ alone (with the 20-bins PCA prescription), the black dashed lines show results using only the 2PCF $\xi$, while the filled red contours represent the combined analysis using the full PCA data vector ($\xi + \zeta_m$). 
We find no improvement in the $S_8$ and $w_0$ parameters estimations when going from $\xi$ alone to $\xi+\zeta_m$. However, the constraining power on $\Omega_m$ and $\sigma_8$ has grown significantly, in around 28\% and 29\%,  respectively. When accounting for the Percival factor for the 2PCF alone (18 bins), which is 1.15,  and for the combined 2PCF+3PCF (38 bins), which is 1.52, the statistical improvement is reduced to 17\% and 18\% for  $\Omega_m$ and $\sigma_8$, respectively.  %Percival38=1.47783 #Percival30=1.32433 #Percival29=1.30736 #Percival18=1.14581

We expect the constraining power and degeneracy directions change with redshift. Hence, we also test galaxy distributions $W_g(\chi)$ centered at $z=1.0$
and $z=2.0$, each one with a width of $150 \, h^{-1} \text{Mpc}$. 
As expected, the analytical model is less accurate at these redshifts than at $z=0.5$. Consequently, for the 2PCF, we are forced to remove more bins, retaining only the 9 (10) largest scales bins for $z=1.0$ ($z=2.0$). Although we also observe larger discrepancies between theory and data for the 3PCF, the corresponding error bars become large enough that the same bins used in the $z=0.5$ analysis pass the filters outlined in \cref{sec:modelvsdata}, and hence these bins are also included in the final data vector. 

In the middle and bottom panels of \cref{fig:tri_zs9} we plot the contours for samples at galaxy sources $z=1.0$ and $z=2.0$, respectively. In these cases, we find significantly weaker constraints. However, the degeneracies between $\Omega_m$ and the other parameters also change. As a byproduct, at these redshifts, $S_8$ and $w_0$ become more tightly constrained. Such improvement in $S_8$ was recently observed in \cite{2025arXiv250303964G}. 

The results for the three galaxy redshifts, considering only the combination of $\xi$ and $\zeta_m$, are shown in \cref{fig:results_allz}, where the tighter constraints at redshift $z = 0.5$ can be more clearly seen. It is worth emphasizing that for the $z = 1.0$ and $z = 2.0$ cases, the number of bins was reduced from the 38 used at $z = 0.5$ to 31 and 30, respectively. This reduction also led to Hartlap factors closer to unity, thereby reducing the penalization of our constraints due to the limited sample of 108 realizations from the full-sky ray-tracing simulations.

\section{Gaussian analytical covariance}\label{sec:GaussianCov}

We have worked with a compressed data vector %consisting of $n=38$ elements derived from the 2PCF and 3PCF. The compression is 
designed to preserve most of the cosmological information while reducing the dimensionality to a level more compatible with the number of available simulations. Ideally, estimating a reliable covariance matrix for such a data vector would require at least an order of magnitude more independent realizations. However, our analysis relies on a set of only 108 full-sky simulations. In this regime, the sample covariance becomes noisy, and its inverse becomes significantly biased. This has been partly, but not completely, mitigated by considering the Hartlap correction and the Percival factor. 
This motivates the development of theoretical models for the 3PCF covariance. In this section, we present an analytical covariance matrix computed assuming the Gaussian approximation. We test its performance by comparing it to the covariances estimated from the Takahashi simulations for single multipoles, where the sample covariance matrix is more reliable. 

Our derivation, under the Limber approximation, follows closely the 3-dimensional case presented in \cite{Slepian:2015qza,Hou:2021ncj}, and is presented in \aaa{\cref{app:cov}}. The covariance matrix between multipoles $\xi_m(\theta_1,\theta_2)$ and $\xi_{m'}(\theta'_1,\theta'_2)$ becomes the rank-4 tensor
\begin{align} \label{Cmm'}
    C_{mm'}(\theta_{1},\theta_{2};\theta'_{1},\theta'_{2}) &= \frac{2\pi}{A} \int_{0}^{\infty} r dr \Big\{ \xi(r) [f_{m,m'}(r;\theta_{1},\theta'_{1}) f_{m,m'}(r;\theta_{2},\theta'_{2}) \nonumber \\ 
    & + f_{m,m'}(r;\theta_{1},\theta'_{2}) f_{m,m'}(r;\theta_{2},\theta'_{1})] + f_{m,m'}(r;\theta_{1},\theta'_{1}) f_{m}(r;\theta_{2}) f_{m'}(r;\theta'_{2}) \nonumber \\ 
    & + f_{m'}(r;\theta'_{1}) f_{m,-m'}(r;\theta_{1},\theta'_{2}) f_{m}(r;\theta_{2}) + f_{m}(r;\theta_{1}) f_{m,m'}(r;\theta_{2},\theta'_{2}) f_{m'}(r;\theta'_{1}) \nonumber \\ 
    & + f_{m}(r;\theta_{1}) f_{m,-m'}(r;\theta_{2},\theta'_{1}) f_{m'}(r;\theta'_{2})  \Big\}\,,
\end{align}
with $\xi$ the 2PCF given in \cref{eq:xi}, and we defined 
\begin{align}
    f_m(r;\theta) &=  \int_0^{\infty} \frac{ \ell d\ell}{2\pi} C_\ell(\ell) J_{m}(\ell \theta) J_{m}(r\ell),
    \label{Ec:f_m} \\
    f_{m,m'}(r;\theta_{i},\theta'_{j}) &= \int_0^{\infty} \frac{ \ell d\ell}{2\pi} C_\ell(\ell) J_{m}(\ell \theta_{i} ) J_{m'}(\ell \theta'_{j}) J_{m+m'}(r\ell),    
\end{align}
with $C_\ell$ the weak lensing convergence angular power spectrum. We add Poissonian noise to the power spectrum by replacing
\begin{equation}
    C_\ell \rightarrow C_\ell + \frac{\sigma_\epsilon^2}{2 \bar{n}_g},
\end{equation}
with $\bar{n}_g$ the number density of galaxies per steradian (sr), and $\sigma_\epsilon$ is the root-mean-square intrinsic ellipticity, that we take equal to 0.3, as it is usual. For the Takahashi simulations used in this work, the galaxy number density is $n_g=5.1\times 10^{-8} \text{ sr}^{-1}$, resulting in a noise term of $6.1 \times 10^{-11}$. Such a small noise value is unrealistic for actual surveys, contributing very little to the $C_\ell$ values at the scales of interest, accounting for only a few percent of the total variance. 

\begin{figure*}
	\begin{center}
  \includegraphics[width=3 in]{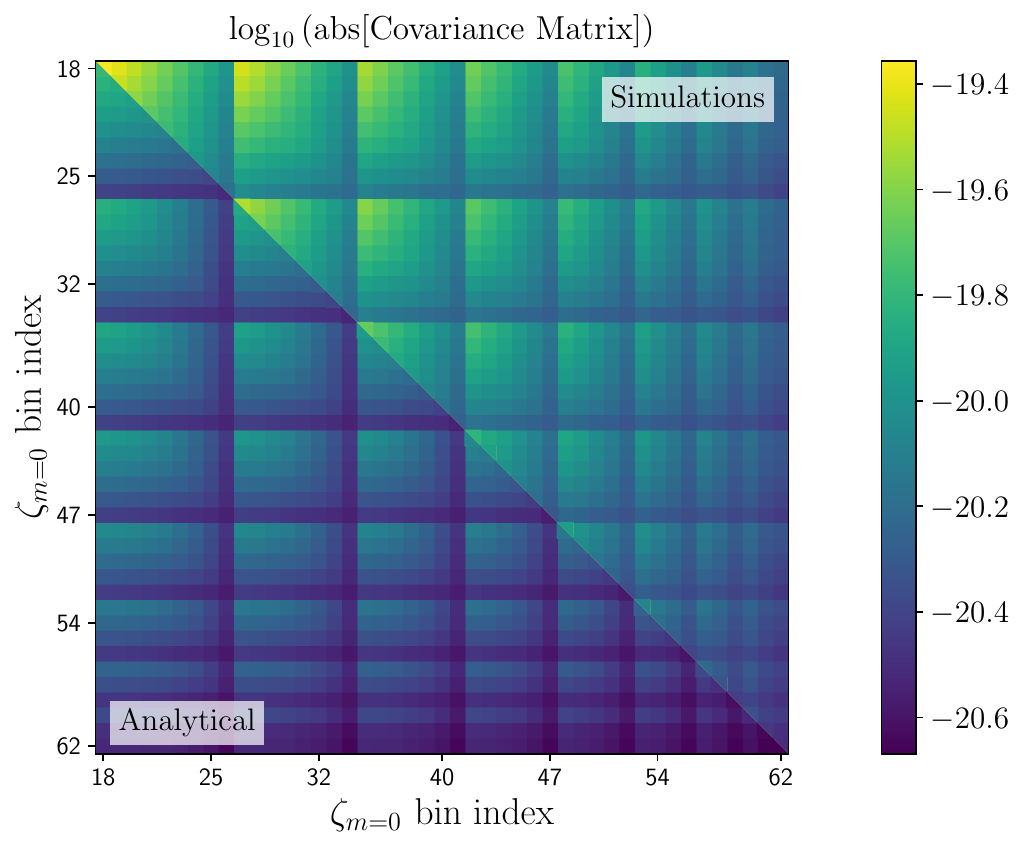}
  \includegraphics[width=3 in]{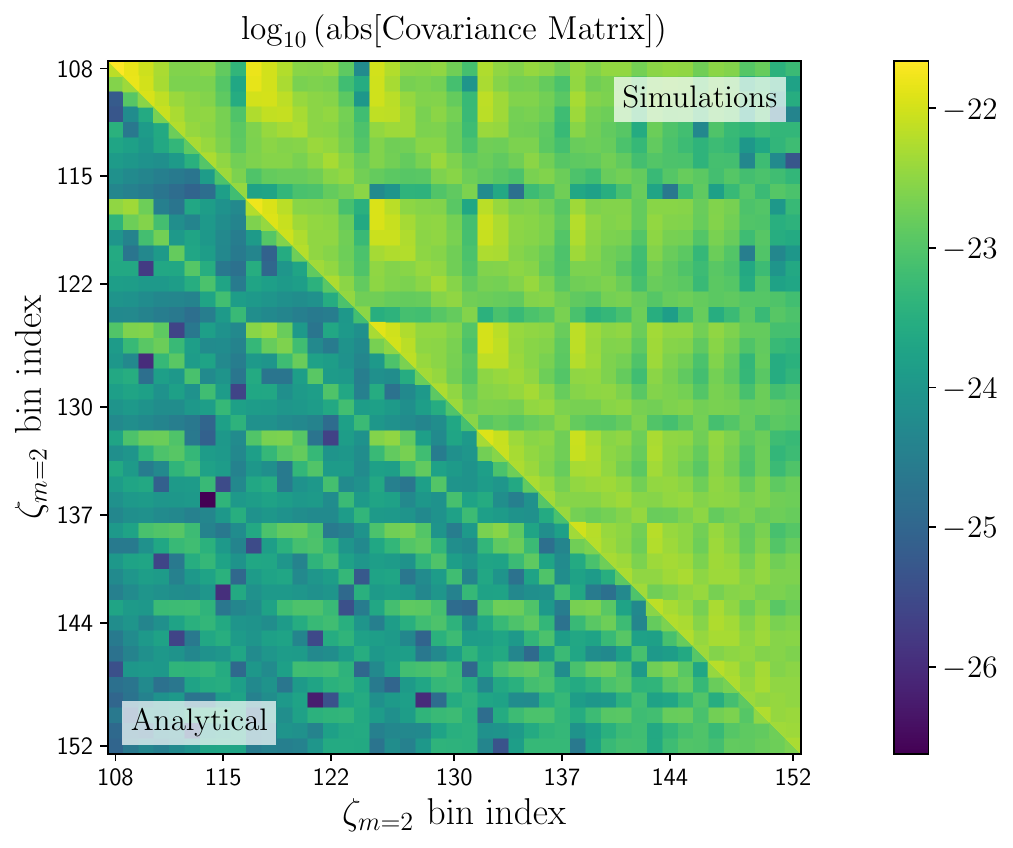}
    \includegraphics[width=3 in]{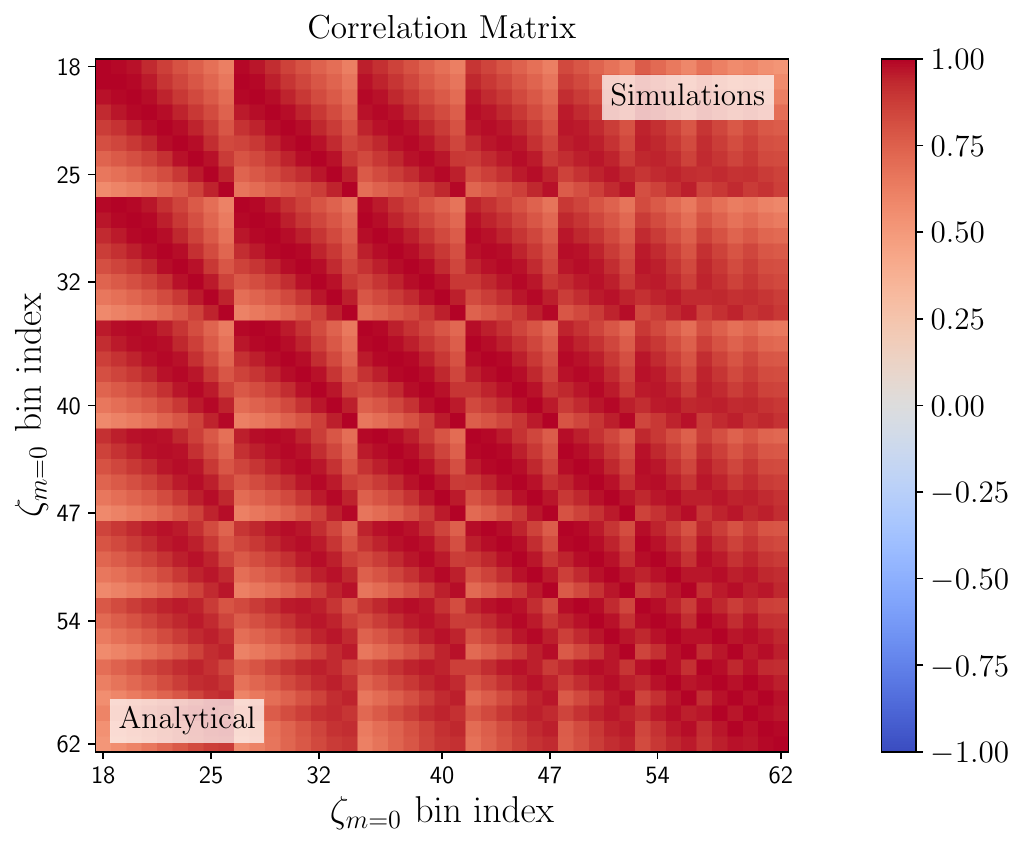}
    \includegraphics[width=3 in]{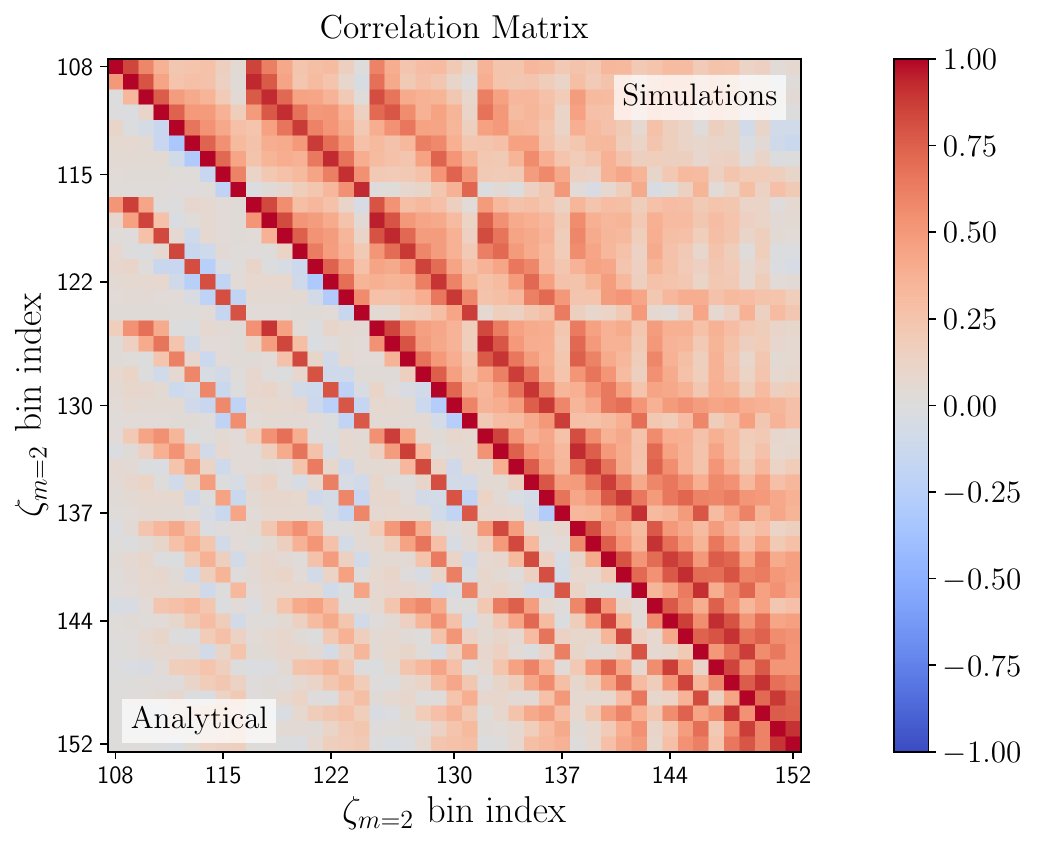}
 \caption{Covariance and correlation matrices for convergence maps at $z=0.5$ obtained from the Takahashi simulations (upper triangles on each panel) and Gaussian theory  (lower triangles). The upper panels shows the covariances for multipoles $(m,m’) = (0,0)$ and $(2,2)$, while the lower panels their corresponding correlation matrices.}  \label{fig:cov_thy_sim}
	\end{center}
\end{figure*}

We have ordered the pairs $(\theta_1,\theta_2)$ as depicted with checkmarks in \cref{fig:zetam_bins}. Hence, we construct the analytical covariances for each multipole $\zeta_m(\text{bin})$. In the top panels of \cref{fig:cov_thy_sim} we present the cases $(m,m')=(0,0)$ and $(m,m')=(2,2)$, that is, the auto-covariances of the monopole and quadrupole, which are the two more constraining multipoles. For an easy comparison, we split each covariance along the diagonal such that in the upper right section we plot the covariance computed from the simulations, while in the lower left the analytical covariance obtained from \cref{Cmm'}. 

For the monopole, the differences between theory and simulations arise mainly as a normalization factor, which changes smoothly from about 0.4 at small angles to 0.8 at large angles, where the Gaussian prescription underestimates the sample covariance. In other words, regions of the covariance matrix corresponding to larger scales tend to yield more similar results between theory and simulations. This is expected, since large scales have undergone less non-linear collapse than the small scales. Despite the Gaussian approximation not being quite accurate in the normalization, the shape of the covariance is remarkably similar for the monopole, this observation is reinforced by observing the correlation matrix in the lower left panel of \cref{fig:cov_thy_sim}, where no difference can be spotted by eye, the differences are indeed quite small ranging from 2\% to 10\%.

For the quadrupole, the differences are more pronounced, not only in normalization factors, which can reach up to a factor of 10, but also in the overall patterns, which are less consistent compared to the monopole case. Further, the upper-right panel shows that noise becomes more significant in the quadrupole. This is consistent with the tendency of the oscillatory plane waves $e^{\text{i}m\phi}$ to cancel out the 3-point signal. However, we attribute the larger discrepancies between simulations and analytical predictions to the more non-Gaussian nature of the quadrupole compared to the monopole. Despite these differences, we also highlight the similarities, particularly evident in the correlation matrix displayed in in the lower left panel of \cref{fig:cov_thy_sim}, showing very similar structures out of the diagonal.

Although this analysis shows that the analytical Gaussian covariance matrices preserve the structure of the sample covariance matrices, the scales do not match: The discrepancy is considerably larger than that observed for galaxy count statistics used in spectroscopic surveys at large scales (e.g. \cite{Hou:2021ncj}). Hence, we do not attempt to obtain forecast from them, which would underestimate the confidence regions. 

The discrepancies  we observe between sample and analytical covariance matrices are expected from early works with the BOSS DR12 CMASS galaxies \cite{Slepian:2015hca}, which opt to compute Gaussian analytical covariances while leaving the survey volume and galaxy number density as free parameters to match sample covariances. In our case, varying the area associated to the analytical covariance matrix helps to adjust the scale of the error, but varying $n_g$ alters drastically the correlation patterns between different bins. Another potential alternative is to use semi-analytical covariance matrices, as in the Rascal method \cite{OConnell:2015src,Philcox:2019xzt}, that uses analytic models to capture the Gaussian (disconnected) pieces of the covariance, dominant on large scales, while incorporating measurements from a limited number of simulations to account for the non-Gaussian part and survey geometry effects on small scales. This hybrid approach has some advantages over using the sample-covariance method and is used in analysis of spectroscopic galaxy surveys by the Dark Energy Science Instrument (DESI) collaboration \cite{Rashkovetskyi:2024eik,2025arXiv250314738D}. %\aa{\cite{Rashkovetskyi:2024eik}+?}.

%\begin{figure*}
%	\begin{center}
%  \includegraphics[width=6 in]{figures/Cov_corr_matrices_quadrupole.pdf}
% \caption{...} \label{fig:Cov_corr_quadrupole}
%	\end{center}
%\end{figure*} 

\section{Conclusions}\label{sec:conclusions}

Statistics that capture non-Gaussian profiles of cosmic fields are considerably useful to constrain cosmological parameters and test theories beyond $\Lambda$CDM. They complement 2-point statistics in different ways. They can break degeneracies between parameters, as well as provide additional information that arises from the non-linear gravitational collapse of structures, which is a process that does not preserve the nearly Gaussian profile of primordial fields. However, higher-order statistics pose several numerical, analytical, and observational challenges. In particular, calculating three-point functions from a sample of $N$ points requires averaging products of fields at three different positions, which has a computational complexity $\mathcal{O}(N^3)$. Early \cite{Szapudi:2004gg,Zheng:2004eh,Pan:2005ym} and recent works \cite{Slepian:2015qza,Slepian:2016weg,Slepian:2016kfz,Philcox:2021bwo,Hou:2021ncj,Sugiyama:2024uqo,Arvizu:2024rlt,Porth:2023dzx} have mitigated this inconvenience by decomposing 3PCFs (and more generally $N$-point functions) in harmonic bases, reducing the computational complexity for each of the coefficients of the expansion (the moments $\zeta_m$) to a time $\mathcal{O}(N^2)$, just as standard 2-point statistics scale. The drawback is that formally we need an infinite set of summary statistics to describe the whole 3PCF. No mathematical proof exists of the convergence of these statistics, nor on how many of these coefficients are necessary to extract most of the cosmological information they contain. In this work, we address this question for the case of the convergence of the galaxy weak lensing signal. By using a Fisher forecast analysis, based on an analytical model of the 3PCF and a covariance matrix extracted from simulations, we show that only the first four multipoles, $\zeta_{m=0,1,2,3}$, are sufficient to extract most of the cosmological information available from the 3PCF, the FoM contributions from multipoles with $m > 3$ decay rapidly. To arrive at this conclusion, we have assumed convergence of the multipoles, which is physically expected, and primordial Gaussianity of fields. Apart from the Fisher matrix information analysis, we have used the DALI method, which accounts for the non-Gaussian nature of the likelihood function. We argue that DALI forecasts are useful for weak lensing analysis due to the non-linear degeneracy between cosmological parameters, particularly between $\sigma_8$ and $\Omega_m$.

Through the covariance and correlation matrices of the 3PCF, we have noted that the different moments are highly redundant. Hence, we perform a PCA that allows us to describe the data vector of the 3PCF multipoles with only 20 bins, out of the initial 423 bins, over a range of scales from 50 to 200 arcmin. For smaller scales, our analytical theory is not reliable. Since we estimate the inverse covariance matrix with a set of 108 simulations, the use of a PCA also became a technical requirement. In our final forecasts, when including Hartlap and Percival factors (which are relevant to our analysis), we find that the 3PCF improves the constraint on $\Omega_m$ and $\sigma_8$ by nearly 20\% compared to constraints when using only the 2PCF. We find no significant improvements for parameters $S_8$ and $w_0$ for redshift $z = 0.5$. However, we also analyze convergence maps at $z = 1.0$ and $z = 2.0$, and despite obtaining weaker overall constraints, we observe considerable improvements in the parameters $S_8$ and $w_0$. This can be attributed to their degeneracy with $\Omega_m$, which is nearly absent at $z = 0$. We conclude that the improved constraining power originates from $\Omega_m$ and is transferred to other parameters through their degeneracies. This effect is most likely driven by the different time-scaling relations between the matter and convergence angular power spectra, compared to those between the matter and convergence angular bispectra.

Finally, following \cite{Slepian:2015qza,Hou:2021ncj}, we construct analytical covariance matrices assuming Gaussianity. We find remarkable similarities between theory and simulations, especially for the monopole $\zeta_0$ auto-correlation. However, we also note that the differences are significant, particularly in their amplitude and scale dependence, with the Gaussian results substantially underestimating the errors. Nevertheless, we consider that this analysis can serve as a basis for hybrid, semi-analytical constructions of covariance matrices based on theory and simulations, such as in the Rascal method, which is now routinely used in 3D galaxy count analyses within the DESI experiment.

The analyses of this work show that the cosmological information of the weak lensing 3PCF is mainly contained within its first few multipoles, effectively reducing its dimensionality. Further, our forecasts suggest that by using the harmonic basis expansion methods and PCA compressions, one can achieve a considerable improvement in parameter constraints at a small computational cost, which can serve present and future photometric surveys, such as DES, Euclid, and LSST.

\acknowledgments

We would like to thank Luis Benet, Gustavo Niz, Mario A. Rodriguez-Meza and Sunao Sugiyama for fruitful discussions. We also thank the MCPCovariances topical team and the HOS topical team of the LSST the Dark Energy Science Collaboration. 

The authors acknowledge the LSST-MX Consortium. % for the management to facilitate their participation in the Vera C. Rubin Observatory.

The authors acknowledge financial support by SECIHTI grant CBF 2023-2024-162. AA and SSN acknowledge financial support by PAPIIT IA101825. The authors also acknowledge support from PAPIIT-UNAM grant No. IG102123: "Laboratorio de Modelos y datos para proyectos de investigación científica: Censos Astrofísicos".

\appendix

\section{Analytical Gaussian covariance matrix derivation}
\label{app:cov}

In this section we consider a scalar field $\kappa$ defined over the sphere with the 3PCF given by \eqref{Ec: Estimator}. The covariance matrix is given by
\begin{align}
 \label{Ec.cov}
    \ve C(\vt_1,\vt_2; \vt'_1, \vt'_2) &= \langle \zeta(\vt_1,\vt_2)\zeta(\vt_1',\vt_2') \rangle  - \langle \zeta(\vt_1,\vt_2) \rangle \langle  \zeta(\vt_1',\vt_2')\rangle \nonumber \\
    &= \langle \zeta(\vt_1,\vt_2)\zeta(\vt_1',\vt_2') \rangle,
\end{align}
where we assume Gaussianity in the second equality. An unbiased estimator for the 3PCF is given by
\begin{equation}
    \hat{\zeta}(\vt_1,\vt_2) = \int \frac{d^{2}\vnu}{A} \kappa(\vnu)\kappa(\vnu + \vt_1)\kappa(\vnu + \vt_2),
\end{equation}
where $A$ is the area that we are considering on the sky. Moving the convergence field to Fourier space\footnote{We utilize the convention for the power spectrum
\begin{equation}
\kappa(\vell) = \int \frac{d^2 \vt}{(2\pi)^2} e^{\text{i}\vell\cdot \vt}\,\kappa(\vt).
\end{equation}
We also use the expansion of 2D plane waves on 1D plane waves
\begin{equation}
 e^{\text{i} \vell \cdot \vt} = \sum_{n=-\infty}^\infty i^n J_n( \ell \theta ) e^{\text{i} n \beta}, \quad \text{with} \quad \cos{\beta} = \hat{\ell} \cdot \hat{\theta} 
 \label{property-angles}
\end{equation}
and the property
\begin{equation}\label{Jnminusn}
    J_{-n} (x) = (-1)^n J_n(x).
\end{equation}
}, assuming Gaussian fields \footnote{In this case, we can make a Wick expansion, such that
\begin{align}
\langle \zeta(\vt_1,\vt_2)\zeta(\vt_1',\vt_2') \rangle &=   \int \frac{d^{2}\vnu}{A} \frac{d^{2}\vnu'}{A} \big\langle \kappa(\vnu)\kappa(\vnu + \vt_1)\kappa(\vnu + \vt_2) \kappa(\vnu')\kappa(\vnu' + \vt'_1)\kappa(\vnu + \vt'_2)\big\rangle \\
&= \int \frac{d^{2}\vnu}{A} \frac{d^{2}\vnu'}{A} \sum_\text{perm}   \langle \kappa(\vx_1)\kappa(\vx_2) \rangle
\langle \kappa(\vx_3)\kappa(\vx_4) \rangle
\langle \kappa(\vx_5)\kappa(\vx_6) \rangle,
\end{align}

where the sum is over all the different products of 2-point correlators that can be formed out of vectors $\vx_i=\vnu,\vnu',\vt_1,\vt'_1,\vt_2,\vt'_2$.
} and performing some integrations with the use of Dirac delta functions
\begin{align} 
\ve C &=  \frac{1}{A} \int \frac{d^2 \vell_1 d^2 \vell_2 d^2 \vell_3}{(2\pi)^{2\times 3}}  (2 \pi)^2 \dD(\vell_1+\vell_2+\vell_3)  C_\ell(\ell_1)C_\ell(\ell_2)C_\ell(\ell_3)
 \nonumber\\ 
 &\quad \times e^{-\text{i}(\vell_1 \cdot \vt_1 + \vell_2 \cdot \vt_2)} 
 \Big\{ e^{-\text{i}( \vell_1 \cdot \vt_1' + \vell_2 \cdot \vt_2')} + e^{-\text{i}( \vell_2 \cdot \vt_1' + \vell_1\cdot \vt_2' )} + e^{-\text{i}( \vell_1 \cdot \vt_1' + \vell_3\cdot \vt_2')} \nonumber\\
 &\quad \qquad \qquad \qquad \quad + e^{-\text{i}( \vell_3 \cdot \vt_1' + \vell_1 \cdot \vt_2' )} + e^{-\text{i}( \vell_2 \cdot \vt_1' + \vell_3\cdot \vt_2' )} + e^{-\text{i}( \vell_3 \cdot \vt_1'+ \vell_2\cdot \vt_2' )} \Big\}, \label{Cov}
\end{align}
with $C_\ell(\ell)$ the angular power spectrum 
\begin{equation}
   (2\pi)^2 \dD(\vell+\vell') C_\ell(\ell) = \langle \kappa(\vell)\kappa(\vell')\rangle.
\end{equation}

Now, we project the covariance matrix onto the harmonic basis as follows,
\begin{align} \label{Cmm1}
 \ve C_{mm'} = \frac{1}{(2\pi)^4}\int d \phi_1  d \phi_2 d \phi_1' d \phi_2' \langle \hat{\zeta}(\vt_1,\vt_2)\hat{\zeta}(\vt_1',\vt_2') \rangle
 e^{-\text{i} m \phi } e^{-\text{i} m' \phi' }
\end{align}
with $\phi = \phi_2 - \phi_1$, and $\phi' = \phi_2' - \phi_1'$. $\phi_{1,2}$ ($\phi_{1,2}'$) is the angle between  $\vt_{1,2}$ ($\vt_{1,2}'$) and a fixed direction $\vhn$.

Let us define
\begin{align}
    I^\text{proj}_m = \frac{1}{(2\pi)^2}\int d \phi_1  d \phi_2 \, e^{-\text{i}(\vell_1 \cdot \vt_1 + \vell_2 \cdot \vt_2)} e^{-\text{i} m \phi },
\end{align}
which, in virtue of \cref{property-angles,Jnminusn}, can be written as
\begin{equation}
    I^\text{proj}_m = (-1)^{m} J_{m}(\ell_{1}\theta_{1}) J_{m}(\ell_{2}\theta_{2}) e^{-\text{i}m\varphi_{21}} = (-1)^{m} \mJ_{m}(\ell_{1},\ell_{2})e^{-\text{i}m\varphi_{21}},
\end{equation}
and the same for $\theta'_{1}$ and $\theta'_{2}$, i.e.
\begin{equation}
    I^\text{proj}_{m'} = (-1)^{ m' } J_{ m'}(\ell_{1}\theta'_{1}) J_{ m'}(\ell_{2}\theta'_{2}) e^{-\text{i} m' \varphi_{21}} = (-1)^{ m' } \mJ_{ m' }(\ell_{1},\ell_{2})e^{-\text{i} m' \varphi_{21}}.
\end{equation}
The second equalities in the above equations define functions $\mJ_{m}$ and $\mJ_{m'}$.

After some manipulations of the projected covariance matrix, we arrive at
\begin{align}
    C_{mm'} %&= 
    % \frac{(-1)^{m+m'}}{A} \int \frac{d^2 \vell_1 d^2 \vell_2 d^2 \vell_3}{(2\pi)^{6}} (2 \pi)^2 \dD(\vell_1+\vell_2+\vell_3)  C(\ell_1)C(\ell_2)C(\ell_3) \nonumber\\  
    % &\times \mJ_{m}(\ell_{1},\ell_{2})e^{-\text{i}m\varphi_{21}} \Big\{ \mJ'_{m'}(\ell_{1},\ell_{2})e^{-\text{i}m'\varphi_{21}} + \mJ'_{m'}(\ell_{2},\ell_{1})e^{-\text{i}m'\varphi_{12}} + \mJ'_{m'}(\ell_{1},\ell_{3})e^{-\text{i}m'\varphi_{31}} \nonumber\\
    % & + \mJ'_{m'}(\ell_{3},\ell_{1})e^{-\text{i}m'\varphi_{13}} + \mJ'_{m'}(\ell_{2},\ell_{3})e^{-\text{i}m'\varphi_{32}} + \mJ'_{m'}(\ell_{3},\ell_{2})e^{-\text{i}m'\varphi_{23}} \Big\}, \nonumber\\
    =& \frac{(-1)^{m+m'}}{A} \int \frac{d^2 \vell_1 d^2 \vell_2 d^2 \vell_3}{(2\pi)^{6}} (2 \pi)^2 \dD(\vell_1+\vell_2+\vell_3)  C_\ell(\ell_1)C_\ell(\ell_2)C_\ell(\ell_3) \mJ_{m}(\ell_{1},\ell_{2}) \nonumber\\
    &\times\Big\{ 
    \quad \!\mJ'_{m'}(\ell_{1},\ell_{2})e^{-\text{i}m\varphi_{21}}e^{-\text{i}m'\varphi_{21}} + \mJ'_{m'}(\ell_{2},\ell_{1})e^{-\text{i}m\varphi_{21}}e^{-\text{i}m'\varphi_{12}}  \nonumber\\
    &\qquad + \mJ'_{m'}(\ell_{1},\ell_{3})e^{-\text{i}m\varphi_{21}}e^{-\text{i}m'\varphi_{31}} + \mJ'_{m'}(\ell_{3},\ell_{1})e^{-\text{i}m\varphi_{21}}e^{-\text{i}m'\varphi_{13}}  \nonumber\\
    &\qquad+ \mJ'_{m'}(\ell_{2},\ell_{3})e^{-\text{i}m\varphi_{21}}e^{-\text{i}m'\varphi_{32}} + \mJ'_{m'}(\ell_{3},\ell_{2})e^{-\text{i}m\varphi_{21}}e^{-\text{i}m'\varphi_{23}} \Big\}.
\end{align}

We now reduce the above equation by performing the angular integrals, $d^{2}\vell_{i} = \ell_{i}d\ell_{i}d\varphi_{i}$. That is, we integrate
\begin{subequations}\label{int_I_fran}
\begin{align}
    I^{\text{symm}}_{{\rm{ang}},ij} &\equiv \int d\varphi_1 d\varphi_2 d\varphi_3 e^{-\text{i} m \varphi_{21}}e^{-\text{i} m' \varphi_{ij}}(2 \pi)^2 \dD(\vell_1+\vell_2+\vell_3) \nonumber\\ 
    & {\rm{for}}\quad (i,j) = (1,2), (2,1)\, , \label{I_symm_fran}\\
    I^{\text{asymm}}_{{\rm{ang}},ij} &\equiv \int d\varphi_1 d\varphi_2 d\varphi_3 e^{-\text{i} m \varphi_{21}}e^{-\text{i} m' \varphi_{ij}}(2 \pi)^2 \dD(\vell_1+\vell_2+\vell_3) \nonumber\\ 
    & {\rm{for}}\quad (i,j)  = (3,2), (2,3), (1,3), (3,1)\, , \label{I_asymm_fran}
\end{align}
\end{subequations}

In order to reduce the above integrals, we carry the following procedure. First, we write
\begin{align}
    (2\pi)^{2}\dD(\vell_1+\vell_2+\vell_3) &= \int d^{2}r e^{\text{i}(\Vec{r}\cdot\vell_{1} + \Vec{r}\cdot\vell_{2} + \Vec{r}\cdot\vell_{3})} \int rdr d\alpha \left[ \sum_{n_{1}=-\infty}^{\infty} i^{n_{1}} J_{n_{1}}(r\ell) e^{\text{i}n_{1}\alpha_{r,\ell_{1}}} \right] \nonumber\\
    & \times \left[ \sum_{n_{2}=-\infty}^{\infty} i^{n_{2}} J_{n_{2}}(r\ell) e^{\text{i}n_{2}\alpha_{r,\ell_{2}}} \right]\left[ \sum_{n_{3}=-\infty}^{\infty} i^{n_{3}} J_{n_{3}}(r\ell) e^{\text{i}n_{3}\alpha_{r,\ell_{3}}} \right].
\end{align}
By defining
\begin{equation}
    \mathcal{R}_{n_{1},n_{2},n_{3}}(\ell_{1},\ell_{2},\ell_{3}) \equiv \int_{0}^{\infty} rdr J_{n_{1}}(r\ell_{1}) J_{n_{2}}(r\ell_{2}) J_{n_{3}}(r\ell_{3}),
\end{equation}
we can write
\begin{equation}
    (2\pi)^{2} \dD(\vell_1+\vell_2+\vell_3) = \sum_{n_{1},n_{2},n_{3}=0}(2\pi)e^{-\text{i}(n_{1}\varphi_{1} + n_{2}\varphi_{2} + n_{3}\varphi_{3})} \mathcal{R}_{n_{1},n_{2},n_{3}}(\ell_{1},\ell_{2},\ell_{3}),
\end{equation}
where we have used the fact that we are working with closed triangles, so $n_{1}+n_{2}+n_{3} = 0$.  After further manipulations we arrive at the identities

\begin{equation}
    I_{\text{ang},21}^{\text{symm}}(\ell_{1},\ell_{2},\ell_{3}) = (2\pi)^{4} (-1)^{m+m'}\int_{0}^{\infty} rdr J_{m+m'}(r\ell_{1})J_{m+m'}(r\ell_{2})J_{0}(r\ell_{3}).
    \label{Ec:I_21}
\end{equation}
\begin{equation}
    I_{\text{ang},31}^{\text{asymm}} %&= 
    = (2\pi)^{4} (-1)^{m+m'} \int_{0}^{\infty} rdr J_{m+m'}(r\ell_{1}) J_{m}(r\ell_{2}) J_{m'}(r\ell_{3}).
    \label{Ec:I_31}
\end{equation}
\begin{equation}
    I_{\text{ang},12}^{\text{symm}}(\ell_{1},\ell_{2},\ell_{3}) = (2\pi)^{4}(-1)^{m-m'} \int_{0}^{\infty} r dr J_{m-m'}(r\ell_{1}) J_{m-m'}(r\ell_{2}) J_{0}(r\ell_{3}),
    \label{Ec:I_12}
\end{equation}
\begin{equation}
    I_{\text{ang},13}^{\text{asymm}}(\ell_{1},\ell_{2},\ell_{3}) = (2\pi)^{4}(-1)^{m} \int_{0}^{\infty} r dr J_{m-m'}(r\ell_{1}) J_{m}(r\ell_{2}) J_{m'}(r\ell_{3}),
\end{equation}
\begin{equation}
    I_{\text{ang},32}^{\text{asymm}}(\ell_{1},\ell_{2},\ell_{3}) = (2\pi)^{4}(-1)^{m} \int_{0}^{\infty} r dr J_{m}(r\ell_{1}) J_{m-m'}(r\ell_{2}) J_{m'}(r\ell_{3}),
\end{equation}
\begin{equation}
    I_{\text{ang},23}^{\text{asymm}}(\ell_{1},\ell_{2},\ell_{3}) = (2\pi)^{4}(-1)^{m+m'} \int_{0}^{\infty} r dr J_{m}(r\ell_{1}) J_{m+m'}(r\ell_{2}) J_{m'}(r\ell_{3}).
    \label{Ec:I_23}
\end{equation}
%We realize that the integrals change to the change of inices, in the following way
%\begin{subequations}
    %\begin{eqnarray}
        %I^{{\rm{symm}}}_{{\rm{ang}},21}  &=& I^{{\rm{symm}}}_{{\rm{ang}},12}\, , \quad {\rm{under}}\quad 12\longleftrightarrow 21\quad {\rm{and}} \quad m^{\prime}\longleftrightarrow -m^{\prime}\, , \\
        %I^{{\rm{symm}}}_{{\rm{ang}},32}  &=& I^{{\rm{symm}}}_{{\rm{ang}},23}\, , \quad {\rm{under}}\quad 23\longleftrightarrow 32\quad {\rm{and}} \quad m^{\prime}\longleftrightarrow -m^{\prime}\, , \\
        %I^{{\rm{symm}}}_{{\rm{ang}},31}  &=& I^{{\rm{symm}}}_{{\rm{ang}},13}\, , \quad {\rm{under}}\quad 13\longleftrightarrow 31\quad {\rm{and}} \quad m^{\prime}\longleftrightarrow -m^{\prime} \, . 
    %\end{eqnarray}
%\end{subequations}

Now,  with the aid of equations (\ref{Ec:I_21}-\ref{Ec:I_23}) we can reduce the covariance matrix and obtain
\begin{align}\label{Cmm}
C_{mm'} &= \frac{(-1)^{m+m'}}{A}
 \int \frac{\ell_1 \,d\ell_1 \ell_2\,d\ell_2 \ell_3 \,d\ell_3}{(2\pi)^{6}}    C_\ell(\ell_1)C_\ell(\ell_2)C_\ell(\ell_3) \nonumber\\  
 &\quad \times 
 \Big\{\quad  \,\mJ_m(\ell_1\, , \ell_2) \mJ'_{m'}(\ell_1\, , \ell_2)\,I^{{\rm{symm}}}_{{\rm{ang}},21}(\ell_1,\ell_2,\ell_3) \nonumber\\
 &\qquad\quad + \mJ_m(\ell_1\, , \ell_2) \mJ'_{m'}(\ell_2\, , \ell_1)\,I^{{\rm{symm}}}_{{\rm{ang}},12}(\ell_1,\ell_2,\ell_3) \nonumber\\
 &\qquad\quad + \mJ_m(\ell_1\, , \ell_2) \mJ'_{m'}(\ell_2\, , \ell_3)\,I^{{\rm{asymm}}}_{{\rm{ang}},32}(\ell_1,\ell_2,\ell_3) \nonumber\\
 &\qquad\quad + \mJ_m(\ell_1\, , \ell_2) \mJ'_{m'}(\ell_3\, , \ell_2)\,I^{{\rm{asymm}}}_{{\rm{ang}},23}(\ell_1,\ell_2,\ell_3) \nonumber\\
 &\qquad\quad + \mJ_m(\ell_1\, , \ell_2) \mJ'_{m'}(\ell_3\, , \ell_1)\,I^{{\rm{asymm}}}_{{\rm{ang}},13}(\ell_1,\ell_2,\ell_3) \nonumber\\
 &\qquad\quad + \mJ_m(\ell_1\, , \ell_2) \mJ'_{m'}(\ell_1\, , \ell_3)\,I^{{\rm{asymm}}}_{{\rm{ang}},31}(\ell_1,\ell_2,\ell_3)
 \,\, \Big\}\, .
\end{align}
After some algebraical manipulations the covariance matrix of \cref{Cmm'} follows.

\section{Impact of the galaxy redshift distribution on the 3PCF}

So far, we have assumed the galaxy distribution to be a shell of width $150\, h^{-1}\text{Mpc}$ around each redshift bin, namely $z=0.5$, 1, and 2. That is, we have adopted $W_g=0.00666$ for $1275<\chi /( h^{-1}\text{Mpc}) <1425  $, and zero otherwise.  
In this appendix, we explore the impact of the galaxy distribution on the 3PCF. For this purpose, we use our C code \texttt{3pt-WL}, which can be provided with an arbitrary function $W_g$ via an ASCII file.

\begin{figure*}
	\begin{center}
  \includegraphics[width=6 in]{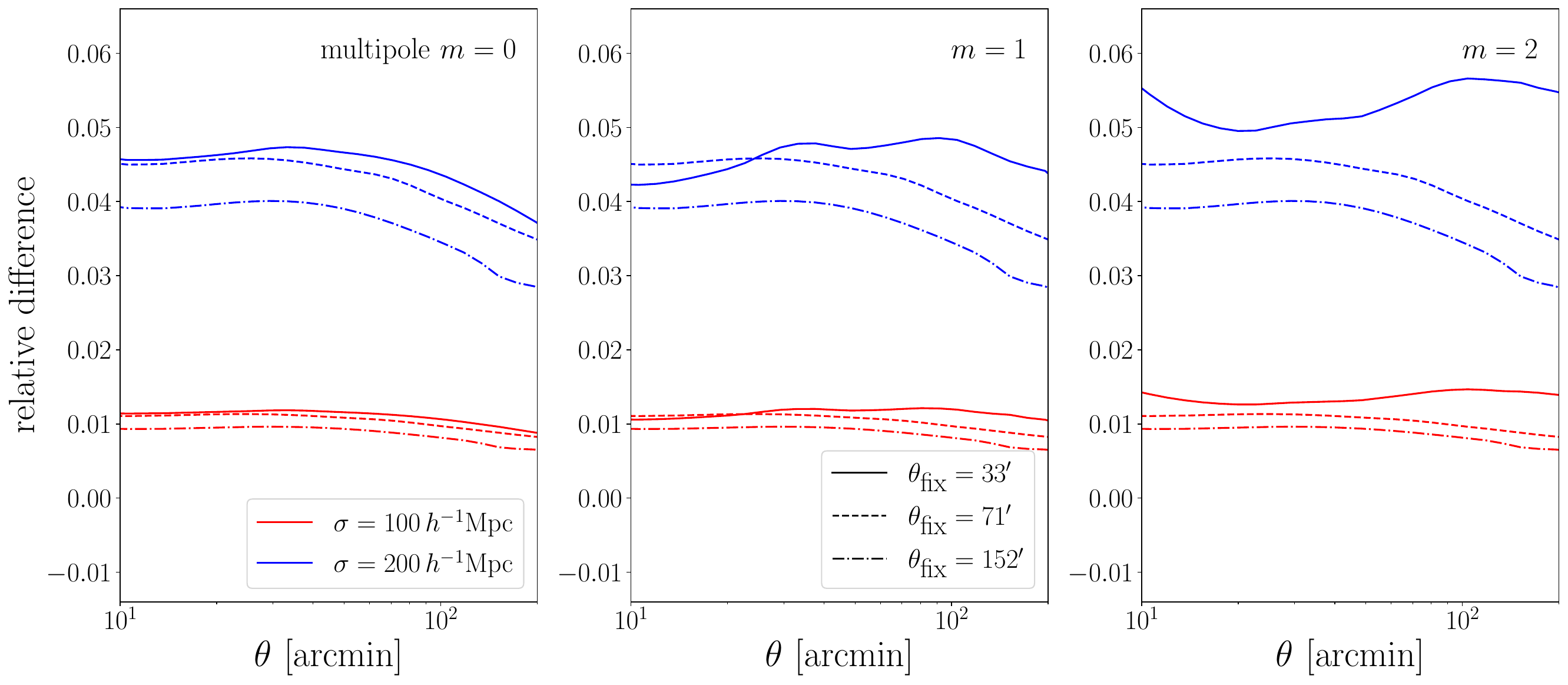}
 \caption{Relative difference, given by \cref{eq:reldiff}, of the galaxy distribution $W_g(\chi)$ used throughout this work, and two Gaussian distributions centered at the same redshift ($z=0.5$), with widths $\sigma=100 \,h^{-1}\text{Mpc}$ (red curves) and $\sigma=200 \,h^{-1}\text{Mpc}$ (blue curves). We fix one of the arguments to $\theta_\text{fix}=33$, 71 and 152 arcmin, shown with solid, dashed and dotdashed lines, respectively. We display the leading multipoles $\zeta_{m=0,1,2}$. }  \label{fig:relativedifferences}
	\end{center}
\end{figure*}

For simplicity, in addition to the layers used so far, we consider two Gaussian distributions, both centered at $\chi=1350 \, h^{-1}\text{Mpc}$ (corresponding to $z=0.508$, under the cosmology assumed in this work), with widths $\sigma=100$ and $200\, h^{-1}\text{Mpc}$. In \cref{fig:relativedifferences}, we show the relative difference with respect to the baseline choice of the thick shell for multipoles $\zeta_0$, $\zeta_1$ and $\zeta_2$. To present this comparison, we fix one of the angles of the 3PCF multipoles to the values $\theta_\text{fix}=33$, 71 and 152  arcmin, and consider the relative difference
\begin{equation} \label{eq:reldiff}
    \frac{\zeta_m^{W_g: \text{\,shell}}(\theta,\theta_\text{fix}) - \zeta_m^{W_g: \text{\,Gaussian}}(\theta,\theta_\text{fix})}{\zeta_m^{W_g: \text{\,shell}}(\theta,\theta_\text{fix})}
\end{equation}
We find that the main effect of adopting different choices of $W_g$ is an overall change in the normalization of the 3PCF multipoles, with only a very small change in their shape. The differences amount to about $1\%$ and $5\%$ for the Gaussian distributions with widths $\sigma = 100$ and $200 \, h^{-1}\text{Mpc}$, respectively. These differences are well within the errors we have considered so far and those expected in observations. Therefore, for realistic surveys, we do not expect our results to be significantly altered by the choice of galaxy distribution.

 \bibliographystyle{JHEP}  % Use the "unsrtnat" BibTeX style for formatting the Bibliography
 \bibliography{refs.bib}

\end{document}